\newif\ifagu
\ifagu
\documentclass{agujournal2019}
  \usepackage{url}
  \usepackage[]{lineno} \linenumbers*[1] 
  \usepackage[inline]{trackchanges}
  \usepackage{soul}
  \draftfalse
  \journalname{JGR: Space Physics}
 \usepackage{amsmath} 
 \let\oldequation\equation
 \let\oldendequation\endequation
  \renewenvironment{equation}
  {\linenomathNonumbers\oldequation}
  {\oldendequation\endlinenomath}
 \let\oldalign\align
 \let\oldendalign\endalign
  \renewenvironment{align}
  {\linenomathNonumbers\oldalign}
  {\oldendalign\endlinenomath}
  \def\maketitle{}
  \def\dateskip{}
  \def\citep{\cite}
\else
\documentclass[12pt]{article}
 \usepackage[margin=1in]{geometry}
 \usepackage{hyperref}
 \usepackage{amsmath}
 \usepackage[sort&compress,numbers]{natbib}
 \def\authors#1{\author{#1}}
 \def\authoraddr#1{}
 \def\authorrunninghead#1{}
 \def\titlerunninghead#1{}
 \date{}
 
 \def\affiliation#1#2{\date{$^{#1}$#2}}
 \def\correspondingauthor#1#2{}
 \def\dateskip{\medskip}
\fi
\usepackage{amssymb}
\usepackage{bm}
\usepackage{graphicx}
\usepackage{tikz}
\usetikzlibrary{external}
\newif\ifextgen
\ifextgen\fi 
\def\wp{w_\parallel}
\def\wr{w_{\parallel R}}
\def\swp{\sqrt{w_\parallel}}
\def\swr{\sqrt{w_{\parallel R}}}
\def\wrn{w_{\parallel R n}}

\begin{document}
\title{Particle Trapping in Axisymmetric Electron Holes}
\authors{I H Hutchinson}
\affiliation{}{Plasma Science and Fusion Center\\ and
Department of Nuclear Science and Engineering,\\ \dateskip
Massachusetts Institute of Technology, Cambridge, MA 02139, USA.}

\maketitle

\begin{abstract}
  Electron orbits are calculated in solitary two-dimensional
  axisymmetric electrostatic potential structures, typical of plasma
  electron holes, in order to establish the conditions for the
  particles to remain trapped.  Analytic calculations of the evolution
  of the parallel energy caused by the perturbing radial electric
  field (breaking magnetic-moment invariance) 
  are shown to agree well with full numerical orbit integration
  Poincar\'e plots. The predominant mechanism of detrapping is
  resonance between the gyro frequency in the parallel magnetic field
  and harmonics of the parallel bounce frequency. A region of
  phase-space adjacent to the trapped-passing boundary in parallel
  energy is generally stochastic because of island overlap of
  different harmonics, but except for very strong radial electric
  field perturbation, more deeply trapped orbits have well-defined
  islands and are permanently confined. A simple universal
  quantitative algorithm is given, and its results plotted as a
  function of magnetic field strength and hole radial scale-length,
  determining the phase space volume available to sustain the electron
  hole by depression of the permanently trapped distribution function.
\end{abstract}

\section{Background}

Electron holes are steady solitary electrostatic positive potential
structures that sustain themselves by an electron density deficit
arising from depressed phase-space density on trapped
orbits \citep{Turikov1984,Schamel1986a,Eliasson2006,Hutchinson2017}. They are
frequently observed in one-dimensional non-linear simulations of
plasma kinetic instabilities \citep{Morse1969,Berk1970,Omura1996,Miyake1998a,Hutchinson2017}, and in observations of space
plasmas \citep{Matsumoto1994,Ergun1998,Bale1998,Mangeney1999,Pickett2008,Andersson2009,Wilson2010,Malaspina2013,Malaspina2014,Vasko2015,Mozer2016,Hutchinson2018b,Mozer2018}. The one-dimensional theory of these hole equilibria is well
established, being a type of BGK mode \citep{Bernstein1957}. However, in multiple
dimensions, both the equilibrium and stability of these self-sustaining
structures is far less well understood. Satellite observations
show that electron holes are generally
three-dimensional \citep{Franz2000,Vasko2017,Holmes2018,Tong2018}, oblate
structures, more extended in the direction perpendicular to the
ambient magnetic field, than parallel, but by an amount that varies
with plasma and hole parameters. Also, analysis and simulation have
shown that initially-one-dimensional holes are subject to
instabilities \citep{Mottez1997,Miyake1998a,Goldman1999,Oppenheim1999,Muschietti2000,Oppenheim2001b,Singh2001,Lu2008,Hutchinson2018,Hutchinson2018a,Hutchinson2019,Hutchinson2019a}
that break them up in the transverse dimension, forming multidimensional remnants.

A significant magnetic field is known theoretically to be necessary
for the existence of multidimensional electron hole equilibria in
non-pathological background electron
distributions \citep{Krasovsky2004,Ng2005,Ng2006}. When the field is
strong enough that the gyro-radius ($\rho$) is very small, the
equilibrium becomes locally one-dimensional \citep{Chen2002,Jovanovic2002a}, with minor corrections to
Poisson's equation to account for any transverse electric field
divergence, but eventually negligible influence on the particle
orbits. At the other extreme, the magnetic field cannot be so weak as
to make the gyro-radius large compared with the hole's transverse
dimensions, otherwise it provides little transverse
confinement. But there is a big parameter range between these two
limits, in which virtually no theory beyond order of magnitude
heuristics has been completed. A high proportion of observed electron
holes have equilibrium parameters lying in this unexplained region,
see for example \cite{Franz2000,Vasko2017,Holmes2018}.

This article presents a first step to carry out rigorous analysis of
multidimensional electron hole equilibria. It adopts a model potential
that is axisymmetric (independent of the angle $\theta$ in a
cylindrical coordinate system) which is a representative subset of
three-dimensional holes. The electron orbits in this equilibrium are
analysed and calculated numerically, to discover which regions of
phase space are permanently trapped; and, in contrast, the regions
that initially possess small enough parallel kinetic energy to be
trapped by the parallel electric field, but evolve soon to become
untrapped, by the transfer of energy from perpendicular gyration. By
time reversal symmetry, equivalent orbits (in equal numbers)
experience evolution of parallel energy from passing to become
trapped. There thus arises a large effective parallel energy diffusion
across the trapped/passing boundary. These detrapping/trapping orbits
cannot sustain depressed electron phase-space density, and so cannot
contribute to maintaining the hole's positive potential, because
the important detrapping occurs usually on a short timescale of a
moderate number of bounces, i.e. approximately of plasma periods.

The present work does not solve the (still unsolved) full problem of
finding a self-consistent equilibrium in which only the velocity
distribution function on the permanently trapped orbits is allowed to
differ from the background distribution. But it does give limits on
what fully self consistent solutions can exist, and indicates what
their distribution functions might look like.

\section{Orbits in axisymmetric electron holes}

We consider the orbits of electrons in a potential $\phi(r,z)$ that is
axisymmetric about the coordinate $z$. This is a 2D problem, meaning
there is just one ignorable coordinate $\theta$. A 2D cartesian
geometry in which one cartesian coordinate is ignorable would give
essentially the same result, and can be considered to be the limit in
which the radius $r$ is large.

\begin{figure}
\ifextgen
  \tikzsetnextfilename{Figure1}
    \centering
  \begin{tikzpicture}\node (image)
    {\ (a)\hskip-2em\includegraphics[width=0.43\hsize]{orbitxy}
    \ (b)\hskip-2em\includegraphics[width=0.56\hsize]{orbitxyz}}
  ;\end{tikzpicture}
\else
\includegraphics[width=\hsize]{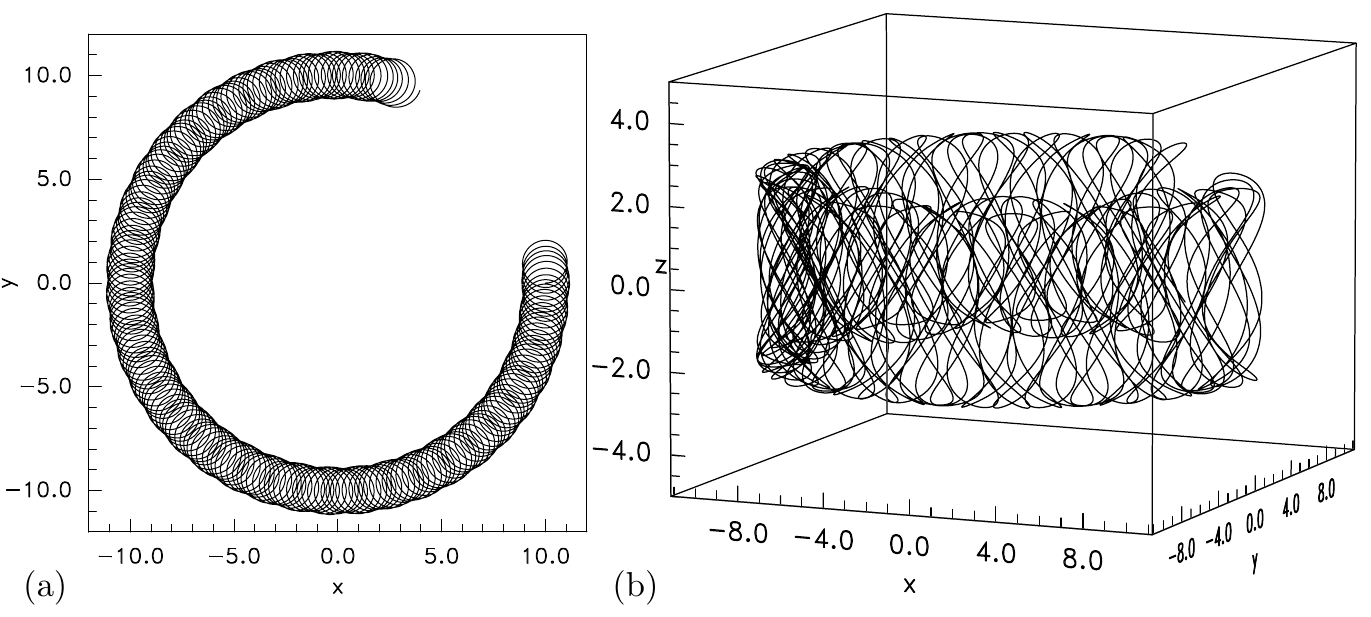}
\fi
\caption{Example of a trapped electron orbit in a model electron-hole
    potential $\phi=\psi\exp([r_0-r]/L_\perp)\,{\rm
      sech}^4(z/4)$, starting at $x=r_0,\ y=0,\ z=0,\ v_x=1,\ v_y=0,\ v_z=1$.
    Viewed (a) in the transverse $x,y$ plane, and (b) in three-dimensions
    showing the bouncing parallel to the magnetic field ($z$)
    direction. (Parameters: $\psi=1$, $\Omega=0.9$, $r_0=10$,
    $1/L_\perp=0.05$.)}
  \label{fig:orbitview}
\end{figure}

In the 1D case where $\phi$ depends only on $z$, there are two exact
constants of the motion which are the total energy $W$ and the
magnetic moment, which in the present uniform magnetic field can be
taken as the perpendicular kinetic energy $W_\perp$. The perpendicular
motion is then entirely decoupled from the parallel and can be
ignored. However, when $\phi$ varies with radius $r$, and a transverse
electric field $E_r$ exists, the $W_\perp$ (magnetic moment)
invariance is broken, and the only strict invariant in addition to $W$
is the canonical angular momentum about the $z$-axis:
$p_\theta=r(v_\theta-\Omega r/2)$, where $\Omega$ is the
gyro-frequency. The effect of $p_\theta$ conservation is mostly to
restrict the range of variation of the orbit's radius to what in the
probe literature are called ``magnetic bottles''
(e.g. \cite{Laframboise1993}). At radii greater than approximately the
gyro-radius $\rho=v_\perp/\Omega$, conservation of $p_\theta$
contributes little to parallel particle dynamics, serving mostly to
localize the orbit in radial position, within approximately one
gyro-radius. Fig.\ \ref{fig:orbitview} illustrates the kind of orbit
that results.

For an electron hole to sustain itself requires a substantial fraction
of the particle orbits to be trapped. These orbits can then
permanently possess a phase-space density ($f$) less than those of
untrapped orbits. Because in a collisionless plasma $f$ is constant
along orbits, the \emph{untrapped} orbits have phase-space density
corresponding to their distribution function at infinity; whereas the
permanently \emph{trapped} orbits have $f$ determined by initial
conditions: the hole formation processes etc. The key question
concerning the existence of a steady solitary electron hole
equilibrium is whether there are enough permanently trapped orbits to
provide a negative electron density perturbation that can sustain the
potential structure self-consistently.

Isotropic multidimensional electron hole equilibria do not exist
because the trapped phase space is then only orbits which have $W<0$,
and in $d$-dimensions this volume $\propto \phi^{d/2}$ is insufficient
when $d>1$ \citep{Krasovsky2004,Ng2005,Ng2006}.  Particle in cell
simulation and drift-orbit analysis show that there exist axisymmetric
2D equilibria, with anisotropic $f(\bm{v})$, when the magnetic field is strong
enough that the gyro-radius ($\rho$) is negligibly small. Essentially
this existence arises because of the adiabatic invariance of $W_\perp$ and
hence $W_\parallel=W-W_\perp$ in the limit of small $\rho$. The
trapped phase-space volume is then larger ($\propto \phi^{1/2}$), requiring only
$W_\parallel <0$ and extending to large positive $W_\perp$.  For the
intermediate case, where $\rho$ is finite, yet transverse
$\phi$-variation exists, the challenge is this. Given that, for finite
magnetic field strength, energy can be exchanged during the orbit
between $W_\parallel$ and $W_\perp$, can one quantify whether and to
what extent the amount exchanged is limited, and an orbit remains
trapped in the $z$-direction ($W_\parallel<0$) even if it has so large
a $W_\perp$ that $W>0$? An earlier attempt on this problem used a more
complicated treatment based on an ``approximate invariant''
\citep{Krasovsky2006} but was not carried through to a full
result. The present paper overcomes the challenge.

Although the calculation will remain as general as possible, we shall
have in mind equipotentials that are oblate: varying faster in the
parallel ($z$) direction than in the transverse ($r$) direction.
For convenience we assume that the $z$-dependence of
$E_r=-\nabla_\perp \phi$ is the same as that of $\phi$, as would be
the case if $\phi$ is of separable form $\phi_r(r)\phi_z(z)$.

We work in units where time is measured in inverse plasma frequencies
($\omega_p=\sqrt{n_ee^2/m_e\epsilon_0}$), length in Debye lengths
($\lambda_{De}=\sqrt{\epsilon_0T_e/e^2n_e}$), and energies (and
potential) in electron temperatures ($T_e$). Thus if primes denote
dimensioned parameters, and unprimed the normalized quantities,
$t=\omega_p't'$, $\bm{x}=\bm{x}'/\lambda_{De}'$, and energy
$W=W'/T_e$. The parameters $T_e$ and $n_e$ are the temperature and
density of the unperturbed electron distribution far from the hole.
In these units the electron mass is eliminated from the equations, and
the electron charge is $q_e=-1$, so the total energy of an electron
can be written $W={1\over2}v^2-\phi$. We shall refer to the
parallel energy as $W_\parallel={1\over2}v_\parallel^2-\phi$ and
perpendicular as
$W_\perp = {1\over2}v_\perp^2={1\over2}(v_r^2+v_\theta^2)$. The
magnetic field strength is represented by the (normalized) cyclotron
frequency $\Omega(=\Omega'/\omega_p')$. The equation of an electron
orbit is then
\begin{equation}
  \label{eq:orbit}
  {d\bm{v}\over dt} = \bm\nabla \phi - \bm{v}\times\Omega\hat{\bm{z}}.
\end{equation}
 
We treat changes in magnetic moment as slow. This is justified if
the transverse electric field (arising from transverse non-uniformity
of $\phi$) is small in the sense that $E_r/\phi\ll 1/\rho$, which may
also be written $L_\perp\gg \rho$, where
$L_\perp \equiv \phi/E_r=\phi/|\partial \phi/\partial r|$ is the
transverse length scale of potential variation. Starting from the
drift limit (which is essentially $\rho/L_\perp\to 0$), we recognize
that the orbit's gyrocenter moves freely along $z$ under the influence
of the parallel electric field, and simultaneously rotates azimuthally
in $\theta$ under the (time-varying) influence of
$\bm{E}_r\times \bm{B}/B^2$.  Trapped orbits (our main focus) bounce
in $z$ and experience an effectively periodic $E_r$ as a
consequence. The mean value of $E_r$ over a period determines the
average azimuthal rotation. The varying component of $E_r$ is the
perturbation responsible for the transfer between perpendicular and
parallel energy. To first order, the fractional transfer of energy in
a bounce period is small. Then during a single period, $z(t)$ can be
approximated as being given by parallel motion with fixed
$W_\parallel$, which is simply the 1D orbit problem. Moreover
$E_r(t)=E_r(r,z(t))=-\partial \phi(r,z(t))/\partial r$ can be
approximated as being at fixed radius $r$ (again provided $\rho$ is
small enough relative to $L_\perp$). The instantaneous energy transfer
rate (recalling that $W=W_\parallel+W_\perp$ is exactly conserved) is
simply the rate of doing work on the electron by $E_r$, namely
\begin{equation}
  \label{eq:workrate}
  {dW_\parallel\over dt} = -{dW_\perp\over dt}= -E_r(t) v_r(t).
\end{equation}
The important velocity component in this equation arises from the gyro
motion of the electron, $v_r=v_\perp \cos(\Omega t)$. When this
equation is integrated over many bounces and gyro-periods, large
excursions in $W_\parallel$ will occur if there is a resonance
between the gyro frequency and a harmonic of the bounce frequency
$\omega_b$. These are the orbits that are liable to lead to
detrapping, because the energy transfer is consistently unidirectional
(between $W_\perp$ and $W_\parallel$) over many bounce periods.

\begin{figure}[htp]
  \centering
  \includegraphics[width=0.6\hsize]{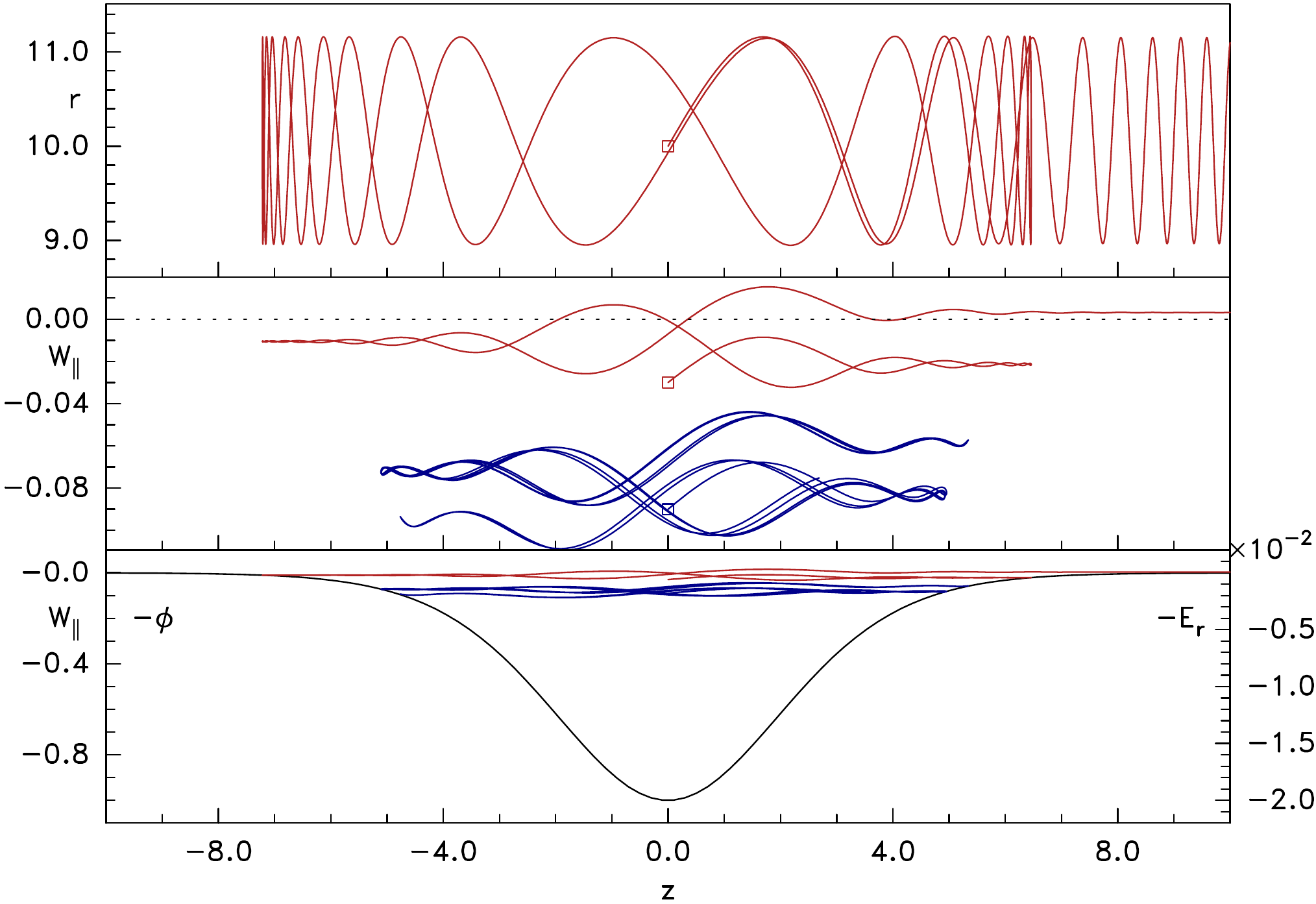}
  \caption{Example of detrapped and permanently trapped
    orbits. Parameters $\psi=1$, $\Omega=1.29$, $W=1$, $E_r=0.02$,
    and initial $W_\parallel=-0.03$(red), $-0.09$(blue) marked with a
    square.
}
  \label{fig:detraporbit1}
\end{figure}

Fig.\ \ref{fig:detraporbit1} illustrates an orbit (red) that quickly
becomes detrapped and a permanently trapped orbit (blue), all as a
function of parallel position.
The top frame shows the track $(r(t),z(t))$ in the $r$-$z$ plane combining
gyro-motion and parallel reflections (the blue orbit is omitted in
this frame for clarity, but has the same radial excursion). The middle and
bottom frames show instead the parallel energy $W_\parallel(z,t)$ and the
bottom shows $-\phi(z)$ and (read on the right scale) $E_r(z)$. The
detrapped orbit has two $z$-reflections before acquiring
$W_\parallel>0$ at the right-hand hole extremity and leaving the hole,
becoming untrapped. The trapped orbit has many more bounces with
excursions in $W_\parallel$ never reaching zero because it lies
somewhat deeper in the potential well.  If near-resonant orbits do not
lead \emph{directly} to detrapping, by raising $W_\parallel$ above
zero, like the red orbit here, then they generally take the form of
``islands'' in the coordinate space $W_\parallel$ versus relative
phase angle (to be explained more fully in a moment). The result then
is that the orbits remain trapped.  The blue orbit and essentially all
in this hole with even more negative initial $W_\parallel$ are of this
type.

Since there are multiple resonances arising from the harmonics of
$\omega_b$, the orbits can become stochastic and the islands broken
up. Very generally, stochasticity begins in Hamiltonian systems
approximately when there is overlap between the separatrices of
adjacent islands \citep{Chirikov1979,Meiss1992}. Indeed, this
principle is called the Chirikov criterion in recognition of its
discoverer who studied resonances between gyro motion and bounces
along the magnetic field in magnetic traps \citep{Chirikov1960}: a
close analog of our current concern.  If an orbit is stochastic, it is
generally \emph{not} permanently trapped, and in principle cannot
contribute to hole sustainment. Our analytic determination of the
orbit trajectories in $W_\parallel$ disregards the radial variation
of $\phi$ and $E_r$: an appropriate approximation for small
gyroradius (but the full orbit integration, also presented, does not).

\section{Islands in energy}

When discussing resonant perturbation islands in a Hamiltonian system,
one generally requires an angle-like coordinate that amounts to the
phase difference between the Hamiltonian orbit and the
perturbation. In the magnetized electron hole with (presumed) uniform
$\Omega$, the phase difference we require is between the gyro motion
(of $v_\perp$ and hence phase of $v_r$) and a perturbing electric
field which we will take as the Fourier component $E_n$ at some
harmonic $n$ of the slowly varying bounce frequency:
$\omega_n=n\omega_b$. The Fourier component has a fixed phase with
respect to the $z$ motion, which we will take as zero when $z=0$. But
because $\omega_b$ varies with $W_\parallel$, the bounce phase has
a variable rate of change with respect to the gyrophase, whose phase
we have taken as zero when $v_r=v_\perp$. We shall write the phase
difference between bounce and gyromotion as $\xi$, so that
\begin{equation}
  \label{eq:phasediff}
  {d\xi\over dt} = \omega_n-\Omega,
\end{equation}
and seek the locus of orbit motion in the plane $\xi,W_\parallel$. Orbits
will then have 
\begin{align}
  \label{eq:orbitave}
  {dW_\parallel\over dt} &=-E_n v_\perp \cos(\omega_nt)\cos(\Omega t)\nonumber\\
&= -E_n v_\perp {\scriptstyle{1\over2}}[\cos(\omega_n-\Omega)t+
\cos(\omega_n+\Omega)t]\nonumber\\
&\simeq-{\scriptstyle{1\over2}}E_nv_\perp\cos\xi, 
\end{align}
and we have dropped the term $\cos(\omega_n+\Omega)t$, because
it is a fast oscillation, compared with the presumed slow evolution of
$\xi=(\omega-\Omega)t$. We shall mention it later.

Let us suppose for initial
illustrative purposes that the variation of $\omega_n$ with
$W_\parallel$ can be approximated linearly
as
\begin{equation}
  \label{eq:omegavW}
{d\xi\over dt} =  \omega_n-\Omega = {d\omega_n\over dW_\parallel} 
(W_\parallel-W_{\parallel R})
=  {d\omega_n\over dW_\parallel} \Delta W_\parallel,
\end{equation}
where $W_{\parallel R}$ is the value of $W_\parallel$ at which exact resonance
occurs ($\omega_n=\Omega$), and that we can take $E_n$ and ${d\omega_n\over
  dW_\parallel}$ to be independent of $W_\parallel$. Then eq.\ (\ref{eq:workrate}) becomes
\begin{equation}
  \label{eq:worbit}
   {d\omega_n\over dW_\parallel} \Delta W_\parallel {dW_\parallel\over
     d\xi}=
   {d\omega_n\over dW_\parallel} {d\Delta W_\parallel^2\over 2d\xi}=
 -E_n v_\perp {1\over2}\cos\xi.
\end{equation}
This expression can be integrated as
\begin{equation}
  \label{eq:wpara2}
   { d\omega_n\over dW_\parallel}\Delta W_\parallel^2 = -E_n v_\perp
  \sin\xi + C.
\end{equation}
This is the island locus, and different values of the integration
constant, $C$, give rise to different trajectories, effectively different
starting $W_\parallel$s. The island's separatrix
corresponds to $C=E_n v_\perp$. The x-point (if ${dW_\parallel\over
  d\omega_b}$  is negative) is at $\xi=-\pi/2$
and the (maximum) half-width of the separatrix (at $\xi=\pi/2$) is then 
\begin{equation}
  \label{eq:sepwidth}
  |\Delta W_\parallel| = \sqrt{2E_n v_\perp \left|{dW_\parallel\over
      d\omega_n}\right|} .
\end{equation}
The island center is at $\xi=\pi/2$,
$C=-E_n v_\perp { dW_\parallel\over d\omega_n}$.

\begin{figure}
  \centering
  \includegraphics[width=0.6\hsize]{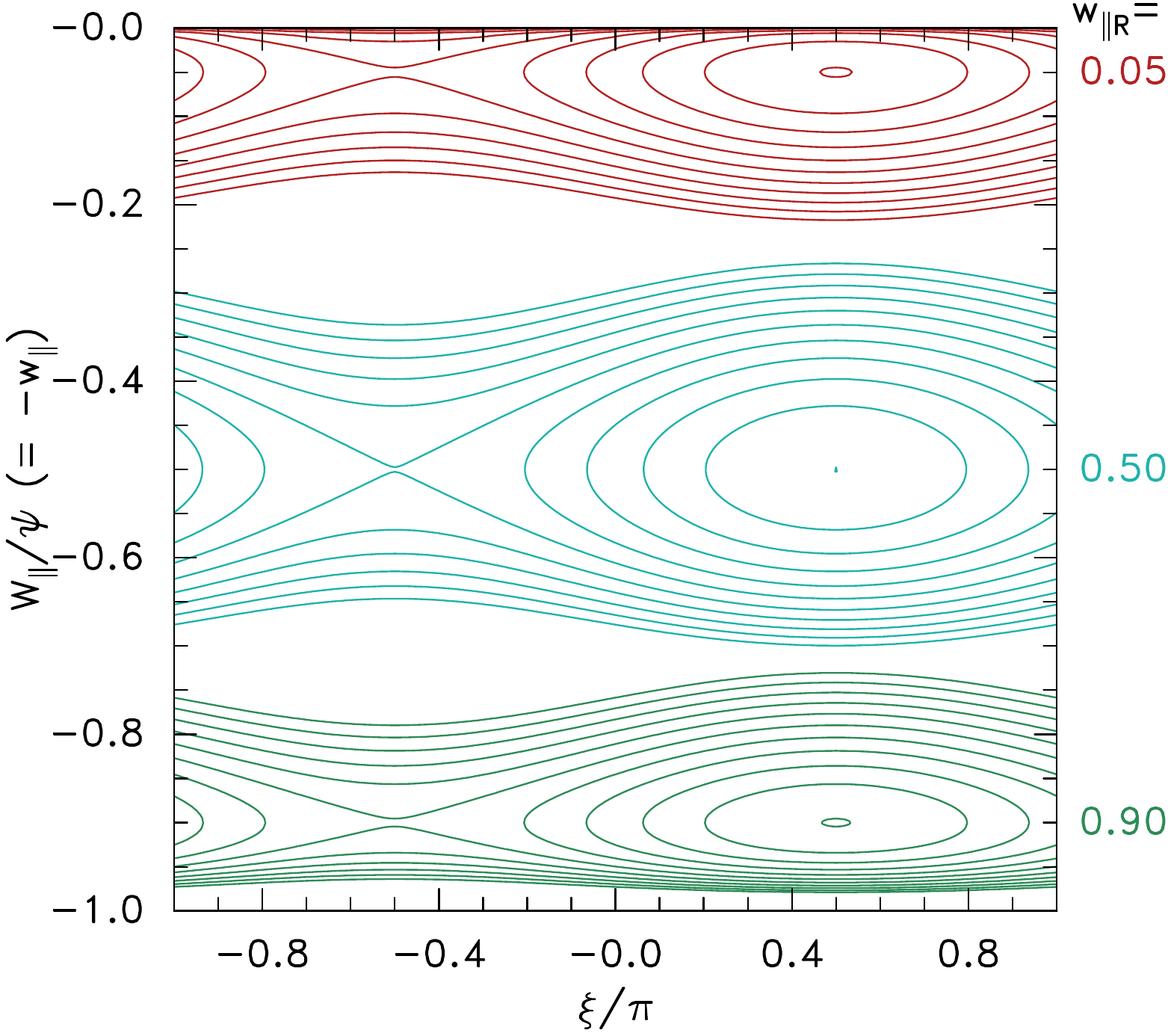}
  \caption{Example energy trajectories for different magnetic field strengths,
  all for the lowest harmonic $n=2$.}
  \label{fig:entraj}
\end{figure}

The result is most usefully plotted as contours of the constant C, in
the $\xi$-$W_\parallel$ plane, which trace the trajectories of the
orbits. Figure \ref{fig:entraj} shows examples from a more elaborate
calculation (and will be explained more fully in section
\ref{illusfig}), but the blue contours in it, centered on
$W_\parallel/\psi=-0.5$, have approximately the shape obtained with the
simple approximations used in this introductory section.

\section{The electric field harmonics}

We must now obtain expressions for $\omega_b$ and $E_n$ as a function
of $W_\parallel$ for a model electron hole. These depend upon the
$z$-profile of the potential, which will be taken as
\begin{equation}
  \label{eq:phiofz}
  \phi(r,z)=\psi(r){\rm sech}^4(z/4).
\end{equation}
This $z$-dependence is what is obtained for 1-D shallow holes whose
trapped distribution is a Maxwellian of negative
temperature \citep{Schamel1979}. More importantly, it falls off at
large distances $\propto \exp(-z)$, which is \emph{required} for
essentially any 1-D Debye shielded potential (at small hole velocity)
that does not have infinite velocity distribution derivative at
$W_\parallel=0$ \citep{Hutchinson2017}. Therefore the results we obtain from this model
potential will apply to shallow trapped orbits for a wide range of
acceptable potential profiles (which is not the case for Gaussian
shaped potentials, often used.)  We approximate the orbit, for the
purpose of determining the $z$-motion, as occurring at fixed $r$
(because of $p_\theta$ conservation), so it is effectively a 1-D
problem in space.

\subsection{Bounce Frequency}
\begin{figure}
  \centering
  \includegraphics[width=0.6\hsize]{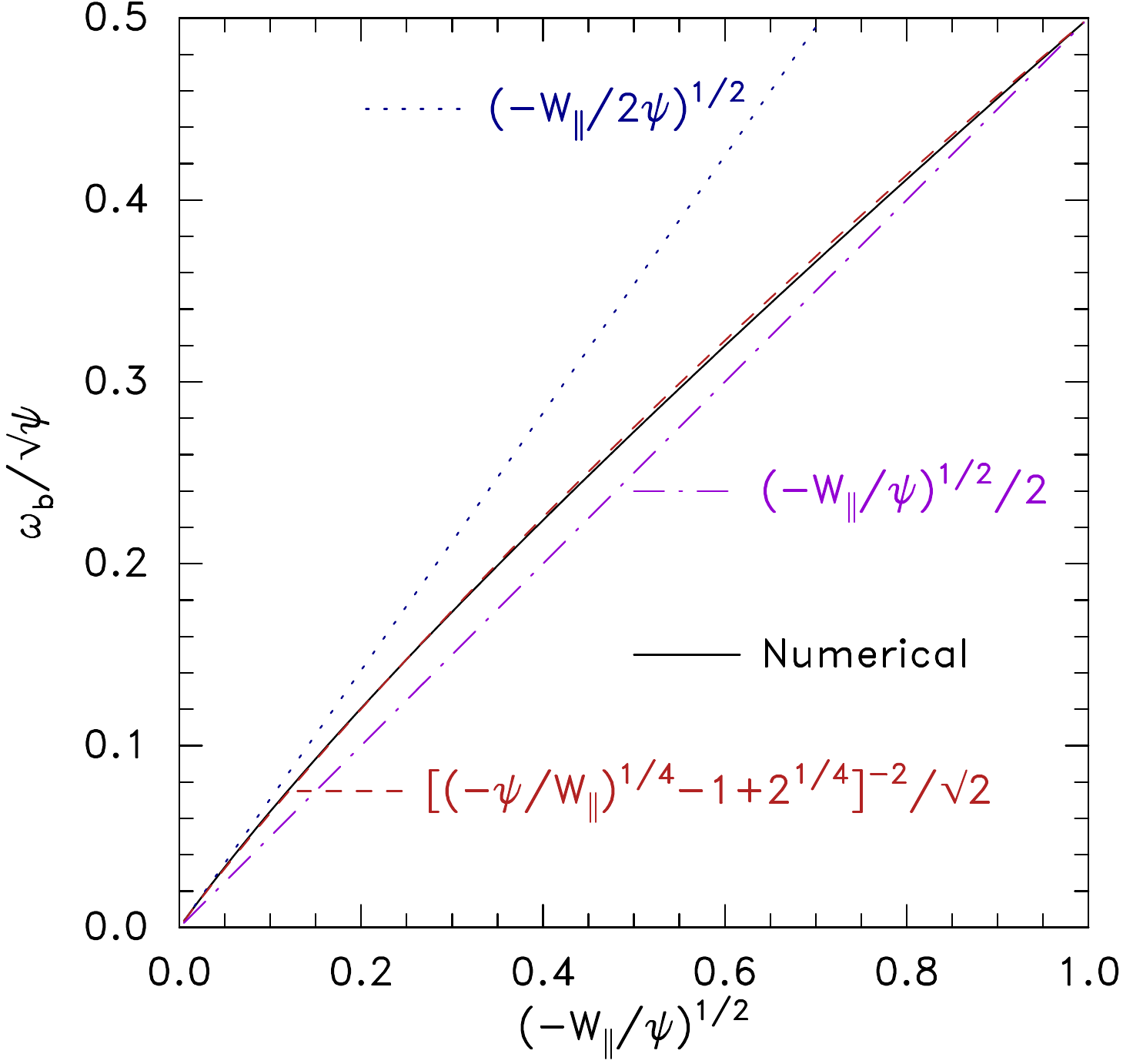}
  \caption{The energy dependence of bounce frequency $\omega_b$ for
    trapped 1-D motion in a potential energy well
    $-\psi\,{\rm sech}^4(z/4)$. Numerical integration gives the solid
    line, and several limits and and approximations are shown.}
  \label{fig:omegabvW}
\end{figure}

It has been shown recently \citep{Hutchinson2019a} that for shallow-trapping
($-W_\parallel\ll \psi$) the (1D) bounce frequency is
$\omega_b\simeq \sqrt{-W_\parallel/2}$ in this potential. Deeply
trapped orbits ($W_\parallel+\psi \ll \psi$) have bounce frequency in
the approxiately parabolic bottom of the potential energy well
$\omega_b=\sqrt{\psi}/2$. This expression is exact for the
${\rm sech}^4$ profile chosen, but the potential shape at the peak can
(unlike the hole wings) be different, so this is a choice of a
particular shape of hole.
It has been found by numerical orbit integration as shown in Fig.\
\ref{fig:omegabvW} that an interpolation of the universal form
\begin{equation}
  \label{eq:omegainterp}
  \omega_b/\sqrt{\psi} = 
\left[(-W_\parallel/\psi)^{-1/4}-1+2^{1/4}\right]^{-2}/\sqrt{2},
\end{equation}
represents the dependence over the entire trapped energy range
extremely well, within approximately the thickness of the line. 
The inverse of this expression is 
\begin{equation}
  \label{eq:omegainverse}
  -W_\parallel/\psi=[(2\omega_b^2/\psi)^{-1/4}+1-2^{1/4} ]^{-4}.
\end{equation}
The
shallow $W_\parallel \to 0$ limit line
$\omega_b= \sqrt{-W_\parallel/2}$ is indicated by the dotted line. 
For
approximate analytic purposes (to avoid the eventual necessity to
evaluate hypergeometric functions) it is adequate to adopt a more
approximate form
\begin{equation}
  \label{eq:omegaapprox}
  \omega_b/\sqrt{\psi} = (-W_\parallel/\psi)^{1/2}/2,
\end{equation}
the dot-dash line with constant slope of $1/2$  in Fig.\ \ref{fig:omegabvW}.

Now we must relate the bounce motion to the time harmonics of
$E_r$. First, observe that for a mirror symmetric potential such as
eq.\ (\ref{eq:phiofz}) the period of the variation of
$E_r=-\partial \phi/\partial r$ with $z$ at constant $r$ is actually
$\pi/\omega_b$, and so only even harmonics $n\omega_b$ are non-zero.
The harmonics $n>2$ arise from the anharmonic motion and the resulting
deviations of $E_r(t)$ from a pure sinusoid.

Let us introduce convenient energy parameter notation involving
positive values normalized to
$\psi$, and (for future use) cyclotron frequency to $\sqrt\psi$ as
\begin{equation}
  \label{eq:scaledtopsi}
  w_\parallel\equiv-W_\parallel/\psi,\qquad w\equiv W/\psi,
  \qquad b\equiv\Omega/\sqrt\psi;
\end{equation}
so trapped orbits have
$\wp$ running from 0 to 1, and the orbits that can become untrapped
have $w>0$. For a given magnetic field value
$b$, and harmonic number $n$, the resonance condition is
$n\omega_b/\sqrt{\psi}=b$,
which gives a resonant parallel energy 
\begin{equation}
  \label{eq:resen}
  \wr=[(2b^2/n^2)^{-1/4}+1-2^{1/4} ]^{-4} 
\end{equation}
corresponding to eq.\ (\ref{eq:omegainverse}).

\subsection{Shallow trapped orbits}
\label{impulsesec}

For shallow-trapped orbits, the $E_r(t)$ has the form of a train of
relatively narrow impulses of width $\sim \tau_t$ and period
$\pi/\omega_b$, which peak briefly as the orbit passes rapidly through
$z\simeq 0$. The orbit spends most of its time near the extrema of the
$z$, where the parallel electric field is very small; and this dwell duration
determines the period \citep{Hutchinson2019a}. The total impulse in a single passage can
be written $A=\int E_r dt$, and its duration is approximately the
potential width divided by the peak speed
[$\tau_t\simeq 8/\sqrt{2(\psi+\wp)}$.] When $\wp/\psi\to 0$,  the integral
$\int E_r(z)dt=\int E_{r0}\,{\rm sech}^4(z/4)dz/v_\parallel(z)$ can be
performed exactly and yields $A=8E_{r0}/\sqrt{2\psi}$. The Fourier
decomposition of $E_r(t)$ then gives the following Fourier mode
amplitudes $E_n$, for even $n$ when $\tau_t\lesssim\pi/\omega_n$ (i.e.\
$\sqrt{-W_\parallel/\psi}=\sqrt{\wp}\lesssim\pi/4n$):
\begin{equation}
  \label{eq:fouriermodes}
  E_n = A {2\omega_b\over \pi} \simeq
  E_{r0}{8\over\pi}\sqrt{-W_\parallel/\psi}
  =E_{r0}{8\over\pi}\sqrt{\wp}.
\end{equation}
We will refer to this as the impulse limit.

\subsection{High Bounce Harmonics}

An alternative perspective of the impulse limit is to note that each
impulse gives an energy change
$\delta \wp= -Av_r=-Av_\perp\cos\xi$, every
$\delta t=\pi/\omega_b$. If $\delta \xi$ and $\delta \wp$
during a single passage through $z=0$ are small, we may approximate the
effect as an average energy rate of change
\begin{equation}
  \label{eq:energychange}
  {d\wp\over dt}={\delta \wp\over\delta t} 
    = -{A v_\perp \omega_b\over \pi}\cos\xi,
\end{equation}
in agreement with eqs.\ (\ref{eq:fouriermodes}), and  
(\ref{eq:orbitave}).

However, if $\Omega\gg\omega_b$, so that only high harmonics of
$\omega_b$ are resonant, the continuum limit is
inappropriate. Moreover, it will always be the case that
$\Omega\gg\omega_b$ near the trapping boundary, $\wp\to 0$, because
$\omega_b\to 0$ there.

When there are many cyclotron periods during one bounce
period, but the cyclotron period is still long compared with the
impulse duration, $\tau_t$ (which does not itself become significantly longer as
$\wp\to 0$), the cyclotron phase ($\xi$) at which each succeeding impulse occurs
becomes effectively random relative to the previous
impulse. So, rather than a systematic continuous flight in the
$(\xi,\wp)$ space, the evolution consists of steps of
virtually random amplitude $\delta \wp$ cosine-distributed
between $\pm Av_\perp$. This represents an effective \emph{diffusion} in
$\wp$ with a diffusion coefficient $\sim(Av_\perp)^2\omega_b/\pi$.
 Moreover, for passing particles, which are
addressed in a recent paper
\cite{Vasko2018} concerning the scattering of passing particles by
successive encounters with different electron holes, one similarly arrives at velocity space diffusion.
The diffusion connects the trapped orbit region $\wp <0$ with
the untrapped region $\wp >0$ across the nominal phase-space
separatrix $\wp=0$, with the result that the distribution
function in this region has only limited gradient $|df/d\wp|$,
and a value approximately equal to the external distribution
$f_\infty$ at $v_\parallel=0$ (in the frame of reference of the hole).
This is one crucial constraint on possible electron-hole equilibria.

\subsection{Deeply trapped orbits}

Orbits that are deeply trapped, having $-W_\parallel/\psi$ a
significant fraction of unity, are not accurately described by the
impulse approximation of the previous section. Instead of being
strongly anharmonic, the $\phi(z)$ is approximately parabolic for
them, and their orbit's $z$-position varies approximately sinusoidally
in time. In the limit $\wp\to 1$, only the lowest Fourier
mode, $n=2$ is important and the higher harmonics become
negligible. Moreover, $E_r$ variation depends on the orbit's $z$-excursion
size; so even for the lowest relevant harmonic $\omega_n=2\omega_b$,
the electric field Fourier amplitude $E_n=E_2$ can become small.

The Taylor expansion of the potential
$\phi(z)=\psi\,{\rm sech}^4(z/4)$ about $z=0$ is
$\phi(z)\simeq \psi[1-z^2/8+7z^4/768+O(z^6)]$, which leads to the
sinusoidal bounce frequency $\omega_b=\sqrt{\psi}/2$ when $z^4$ and
higher terms are dropped. The presumed similar radial electric
field likewise has $E_r(z)\simeq E_{r0}(1-z^2/8)$, of which the time varying
part is only the second term. For given parallel energy,
$\wp$ the amplitude $z_{\max}$ of the $z$-oscillation satisfies
$z_{max}^2/8=(1-\wp)$, and
$
E_r(t)=E_{r0}[1-z_{max}^2\sin^2(\omega_bt)/8]=E_{r0}[(1-z_{max}^2/16)+z_{max}^2\cos(2\omega_bt)/16]$
then yields
\begin{equation}
  \label{eq:E2}
  E_2=z_{max}^2E_{r0}/16=E_{r0}(1-\wp)/2.
\end{equation}
This dependence on $(1-\wp)$ replaces the
$\sqrt{-\wp}$ dependence of eq.\
(\ref{eq:fouriermodes}). 

\subsection{Interpolated $E_n$ expression}

\begin{figure}
  \centering
  \includegraphics[width=0.5\hsize]{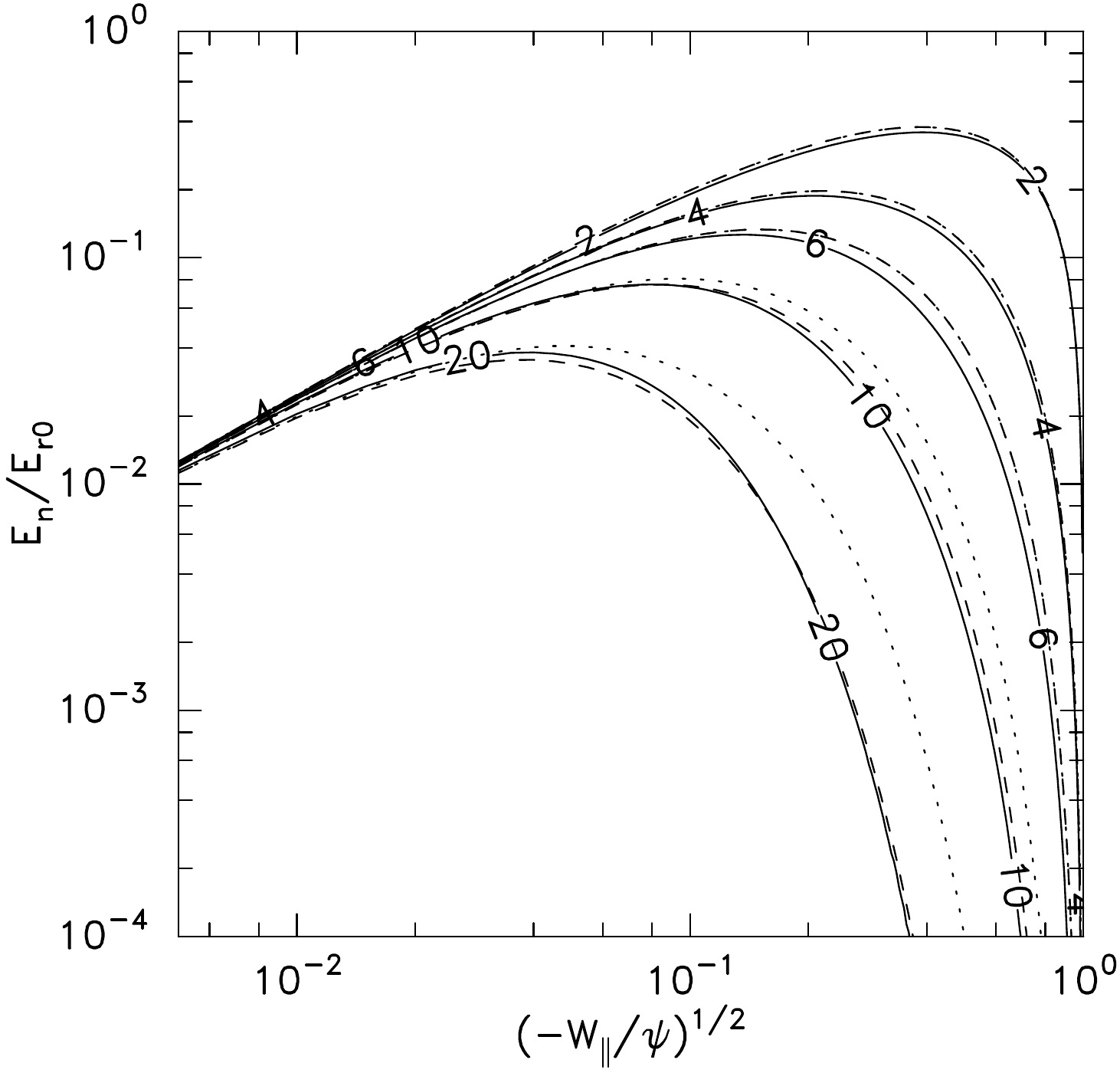}
  \caption{Comparison between numerically integrated Fourier
    coefficients for a ${\rm sech}^4(z/4)$ potential variation (solid
    lines) and the interpolation eq.\ (\ref{eq:Enint}) with $m=n/2$
    (dotted) or using eq.\ (\ref{eq:adhocadj})  (dashed lines),
    for different harmonic number (line labels).}
  \label{fig:Enint}
\end{figure}

It is helpful to have an approximate analytic interpolation for the Fourier
harmonics $E_n$ that spans the entire range $0<\wp<1$. Observe
that $1-\wp=(1-\sqrt{\wp})(1+\sqrt{\wp})$; so
an alternative expression to eq.\ (\ref{eq:E2}), which is equally
valid in the limit $\wp\to 1$, is
$E_2=E_{r0}(1-\sqrt{\wp})$. Realize also that the higher
harmonics, $n=4,6,\dots$, arise from correspondingly higher order
terms in the Taylor expansion of $\phi(z)$, and that therefore, as 
$\wp\to 1$, $E_n$ will become proportional to correspondingly
higher powers:
$(1-\sqrt{\wp})^{n/2}$. Consider then the following proposed interpolation
between the two limits of $\wp$:
\begin{equation}
  \label{eq:Enint}
  E_n= E_{r0} \left[ {n/2\over(1-\sqrt{\wp})^m} + {\pi\over8}{1\over\sqrt{\wp}}\right]^{-1},
\end{equation}
where for most purposes $m=n/2$.
The first term predominates as $\wp\to1$, and the second as
$\wp\to0$. In their respective limits, these two terms give
the correct values for $E_2$, in agreement with eqs.\
\ref{eq:fouriermodes} and \ref{eq:E2}. In the $\wp\to1$ limit,
the higher harmonics have appropriate scaling with
$1-\sqrt{\wp}$. Their numerator $n/2$ has not been
derived; and, more crucially, neither has the inverse form of the
interpolation. Nevertheless, a comparison between this expression and
numerical calculation of the Fourier harmonics, shows quite good
agreement, as can be seen in Fig.\ \ref{fig:Enint}.
This agreement is sufficient for many purposes, but some moderate
discrepancies remain especially at high $n$. They are significantly
reduced if an ad hoc adjustment is made by substituting
\begin{equation}
  \label{eq:adhocadj}
  m=\mathtt{nint}(n/2+\mathtt{max}(n/2-3.3,0)*0.75)  
\end{equation}
(instead of
$m=n/2$) into eq.\ (\ref{eq:Enint}). The adjustment benefits from
retaining convenient integrability.

\section{Solving the $\wp$ trajectories}

\subsection{Analytic calculation}

Using the approximate expression (\ref{eq:omegaapprox}) for $\omega_b$
giving $\omega_n=(n/2)\sqrt{\psi \wp}$, the energy trajectory
equation (\ref{eq:orbitave}) ignoring the fast $\omega_n+\Omega$ term becomes
\begin{align}
  \label{eq:combinedtraj}
  {d\wp \over dt}
  = (\omega_n-\Omega){d\wp \over d\xi}
  &=  {n\sqrt{\psi}\over 2} (\sqrt{\wp}-\sqrt{\wr }){d\wp \over d\xi}
\nonumber\\
  &={1\over2}(E_n/\psi) v_\perp \cos\xi,
\end{align}
where $\wr$ is the resonant parallel energy at which
$\omega_n=\Omega$. Substituting the interpolation for $E_n$ from
eq. (\ref{eq:Enint}), and $v_\perp=\sqrt{2(w+\wp)\psi}$, it can be written
\begin{align}
  \label{eq:combtraj1}
  n{\sqrt{\wp}-\sqrt{\wr }\over\sqrt{2} \sqrt{w+\wp}}
   \left[{n/2\over (1-\sqrt{\wp})^m} + 
     {\pi\over8}{1\over\sqrt{\wp}}\right]
   {d\wp \over d\xi}\nonumber\\
  =(E_{r0}/\psi) \cos\xi.
\end{align}
This equation can be integrated analytically in terms of elementary
functions to obtain
\begin{equation}
  \label{eq:w2solution}
  F_n(\wp,w,\wr ) -(E_{r0}/\psi)\sin\xi=const.,
\end{equation}
where for each $n=2,4,6,\dots$, $F_n$ is a fairly complicated
algebraic expression detailed in the appendix.  For chosen total
energy, magnetic field strength, and perturbing field (i.e.\ $w$,
$\wr $, and $E_{r0}/\psi$) the trajectories can most easily be plotted
as contours of the left hand side expression,
$F_n-(E_{r0}/\psi)\sin\xi$, in the plane $(\xi,\wp)$. In these
calculations it improves accuracy to use the more accurate equation
(\ref{eq:resen}) for $\wr$ in terms of $b$, in $F_n$; and we adopt
this practice forthwith, ignoring the minor inconsistency.

\label{illusfig}
In Fig.\ \ref{fig:entraj} are shown examples of the energy
trajectories for $w=1$, $E_r/\psi=0.01$ $\psi=1$, and three values of the
magnetic field strength, and hence of the resonance energy $\wr $ for
the lowest harmonic $n=2$.  
The perturbing field is quite strong and we can see
that the trajectories near the top or bottom of the potential energy
well (i.e.\ near to $-\wp =$ 0 or -1) are compressed asymmetrically at
those limits because of the form of $F_n$. For an energy away from
those limits, the contours are approximately symmetric about the
resonant energy. If the magnetic field strength is big enough that
$\Omega^2/\psi >1$, then this $n=2$ resonance does not exist.

\begin{figure}
\ifextgen
  \tikzsetnextfilename{Figure6}
    \begin{tikzpicture}\node (image)
  {\noindent
  \includegraphics[width=0.491\hsize]{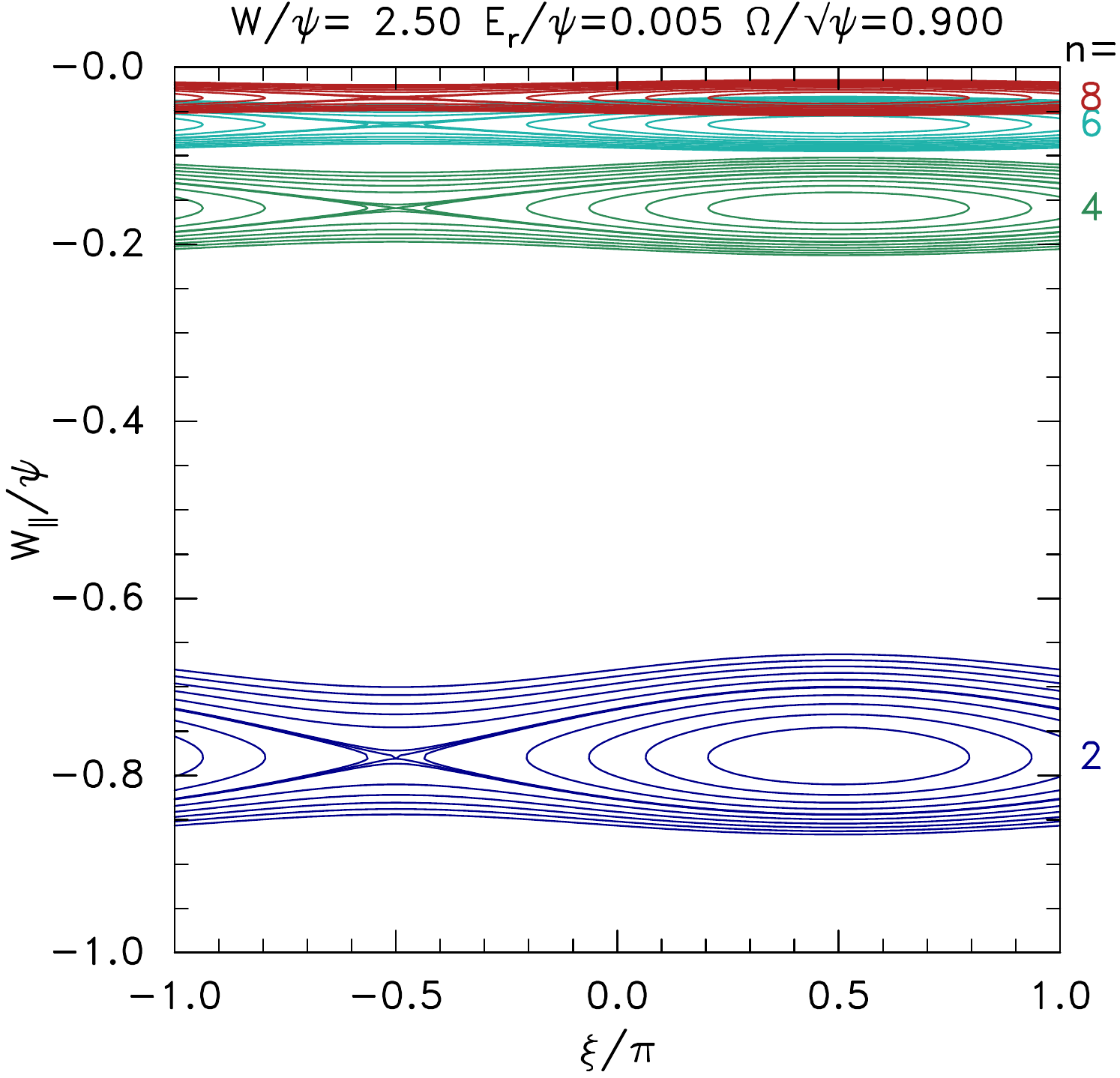}\hskip-2em (a)\hskip1em 
  \includegraphics[width=0.488\hsize]{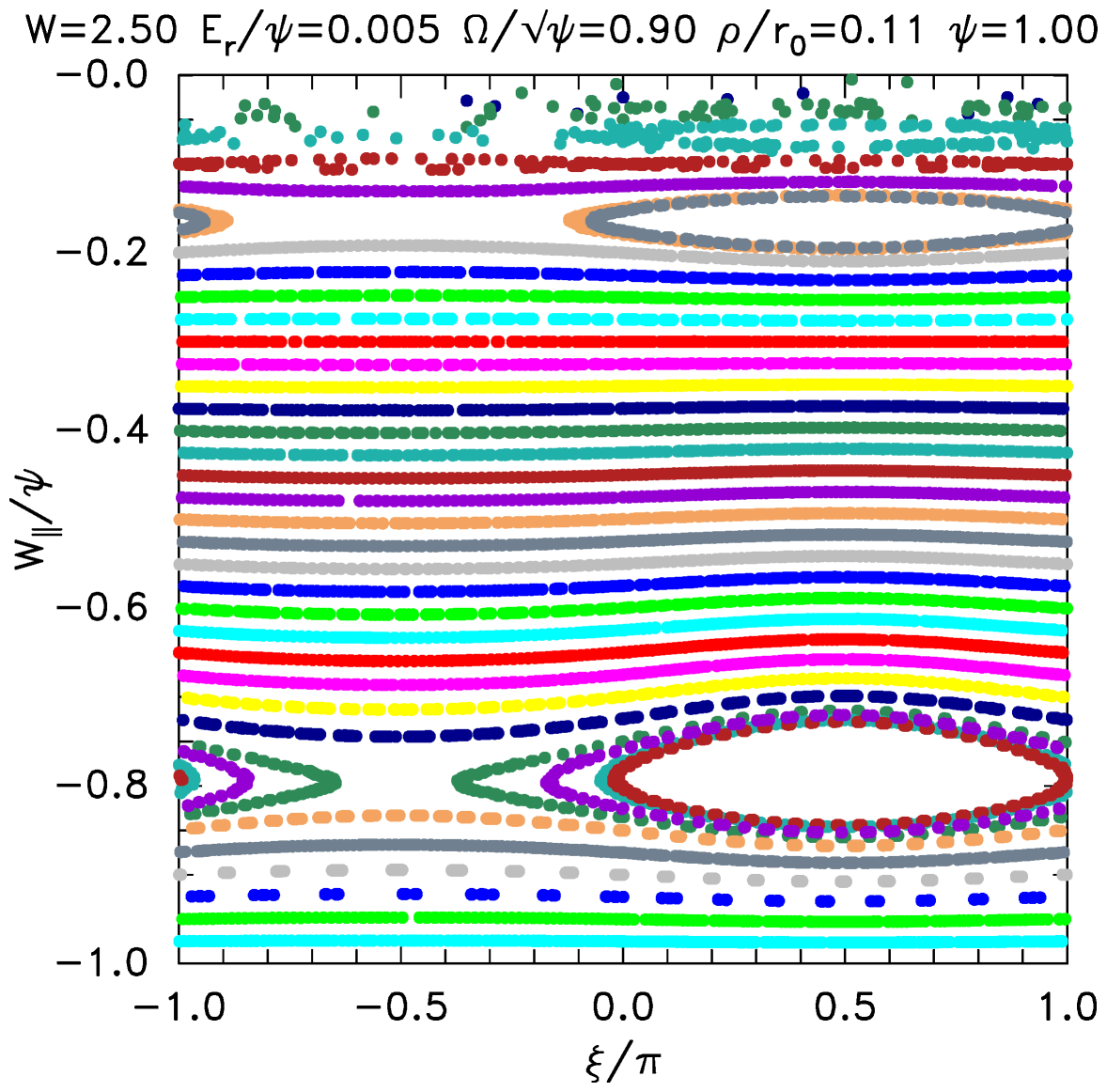}\hskip-2em (b)\phantom{M}}
;\end{tikzpicture}
\else
\includegraphics[width=\hsize]{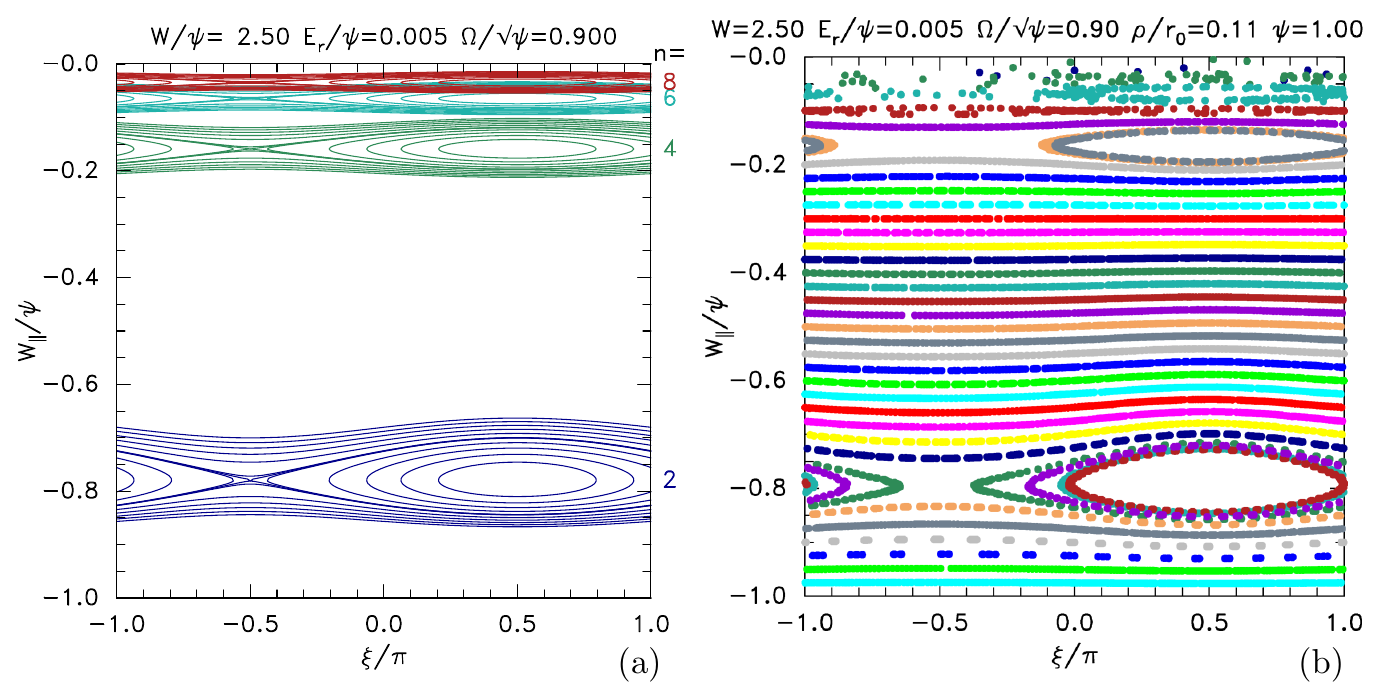}
\fi
  \caption{(a) Example analytic energy trajectories for different harmonics
    ($n$), and fixed magnetic field strength. (b) Poincar\'e plot of
    the corresponding numerically integrated orbit.}
  \label{fig:traj905}
\end{figure}

Fig.\ \ref{fig:traj905}(a), instead shows trajectories for fixed
magnetic field, and hence fixed $n=2$ resonance frequency,
($b=\Omega/\sqrt{\psi}=\sqrt{0.8}$, $E_{r0}/\psi=0.005$), but for
harmonic numbers $n=2,4,6,8,\dots$. The resonance energy is
$\wr =[(2b^2/n^2)^{-1/4}+1-2^{1/4} ]^{-4}$.  The
higher harmonics bunch together near the top of the potential energy
well, corresponding to low bounce frequency. And in fact the $n=6$ and
$n=8$, islands overlap: indicating that this region of energy has
stochastic orbits and so the orbits there are not permanently
trapped. Lower in the well, no overlap occurs with the $n=2$ island;
so orbits there are permanently trapped.

\subsection{Numerical orbits: Poincar\'e Plots}

In order to verify the analytic calculation and to show what happens
when its applicable parameter limits are exceeded, it is helpful to
perform a numerical integration of the trapped orbits. The full
(non-relativistic) equations of motion for the model potential have
been implemented in cylindrical coordinates using a 4th order
Runge-Kutta numerical scheme with timestep chosen short enough that
the (known) conservation of $W$ and $p_\theta$ are reproduced for long
orbits to no worse than 10 times machine precision. This is observed
to require $\Omega.dt\lesssim 0.05$. Figures \ref{fig:orbitview} and
\ref{fig:detraporbit1} are 
examples of orbits so calculated.

Poincar\'e plots of the energy trajectories for such orbits are
obtained by collecting values of $W_\parallel$ and the phase of $v_r$
(i.e.\ \verb!atan2!$(v_\theta,v_r)$) at successive instants when the
orbit passes through $z=0$ (at which the orbit bounce phase is zero or
$\pi$ and the phase of $E_n$ is zero for all even $n$). The phase
difference, $\xi$, thus equals the phase of $v_r$. We place a point at
each of the corresponding positions in $\xi,W_\parallel$ space. We
also, for convenience, start all orbits at $z=0$ and with
$v_\theta=0$, $v_r$ positive: $\xi=0$. We abandon as escaped any
orbits that acquire positive $W_\parallel$ or pass beyond $|z|=20$.  A
technical subtlety is that it is most appropriate to use for
$W_\parallel=v_z^2/2-\phi$, not the value of $\phi$ at the orbit, but
rather the value of $\phi$ at the \emph{gyrocenter}, which gives
significantly smaller oscillatory excursions of $W_\parallel$. It
therefore more effectively suppresses the $\omega_n+\Omega$ term and
expresses the approximate magnetic moment conservation.

Fig.\ \ref{fig:traj905}(b) shows an example of a Poincar\'e plot,
alongside its analytic energy trajectories \ref{fig:traj905}(a). The
agreement is excellent. Orbits are initialized at equally spaced
$W_\parallel$ values. Of course, they cannot trace island contours
well inside their separatrices where the island does not extend past
$\xi=0$. The position and $W_\parallel$-width of the $n=2$ island
agree quantitatively very well between (a) and (b). And the $n=4$ and
$n=6$ islands are also readily seen at their expected
positions. Between the islands, the Poincar\'e points trace the open
contours. Above the position of the $n=6$ island
($W_\parallel/\psi\ge-0.05$) and near its x-point the plot shows
rather incoherent scatter of the points. Orbits above this energy are
stochastic, and terminate after some tens of bounces by leaving the
domain. Again, this agrees well with the analytic observation of
overlap between $n=6$ and $8$, but not between $n=4$ and $6$ islands.
\begin{figure}
\ifextgen
  \tikzsetnextfilename{Figure7}
      \begin{tikzpicture}\node (image)
  {\noindent
  \includegraphics[width=0.491\hsize]{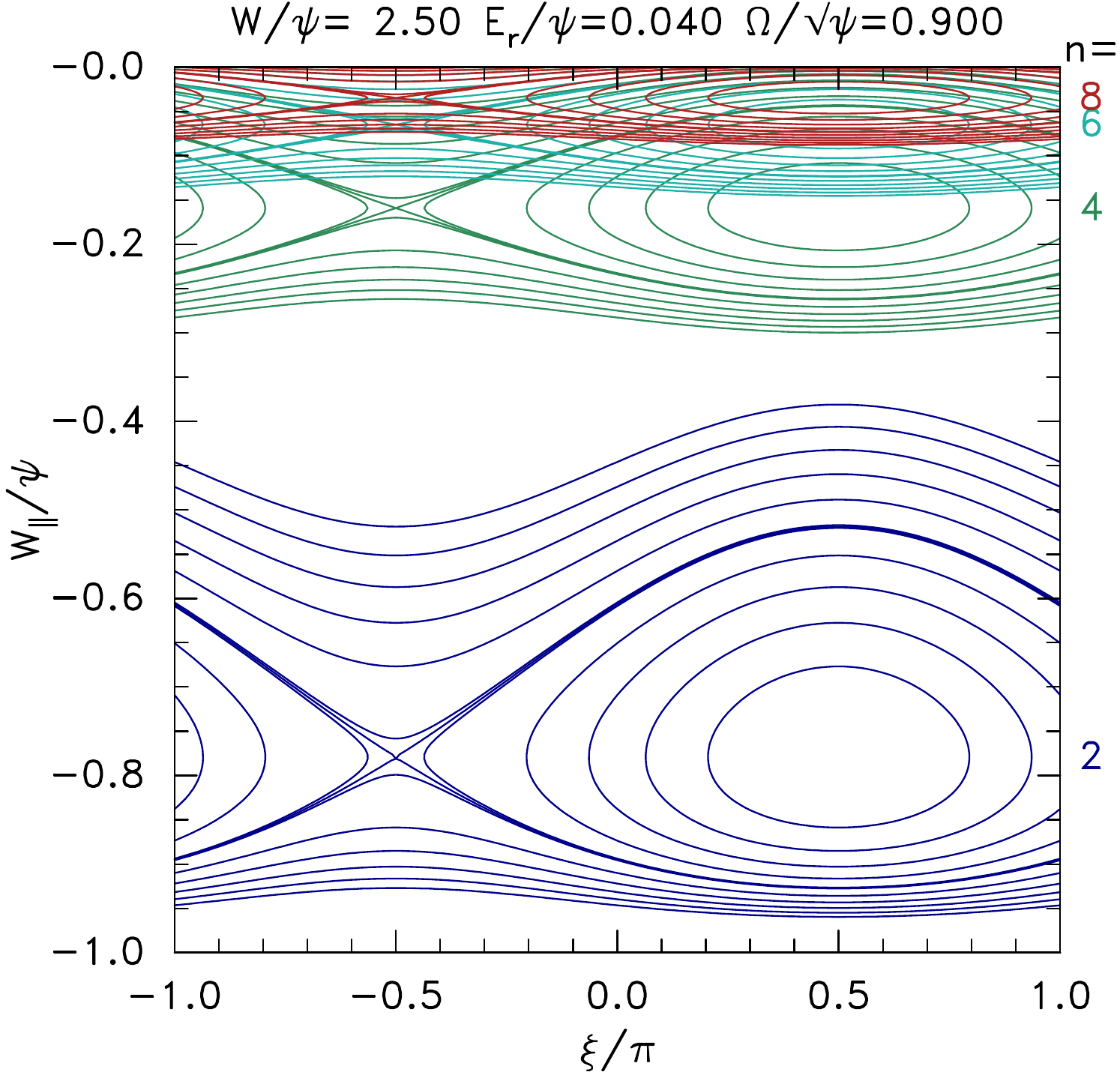}\hskip-2em (a)\hskip1em 
  \includegraphics[width=0.488\hsize]{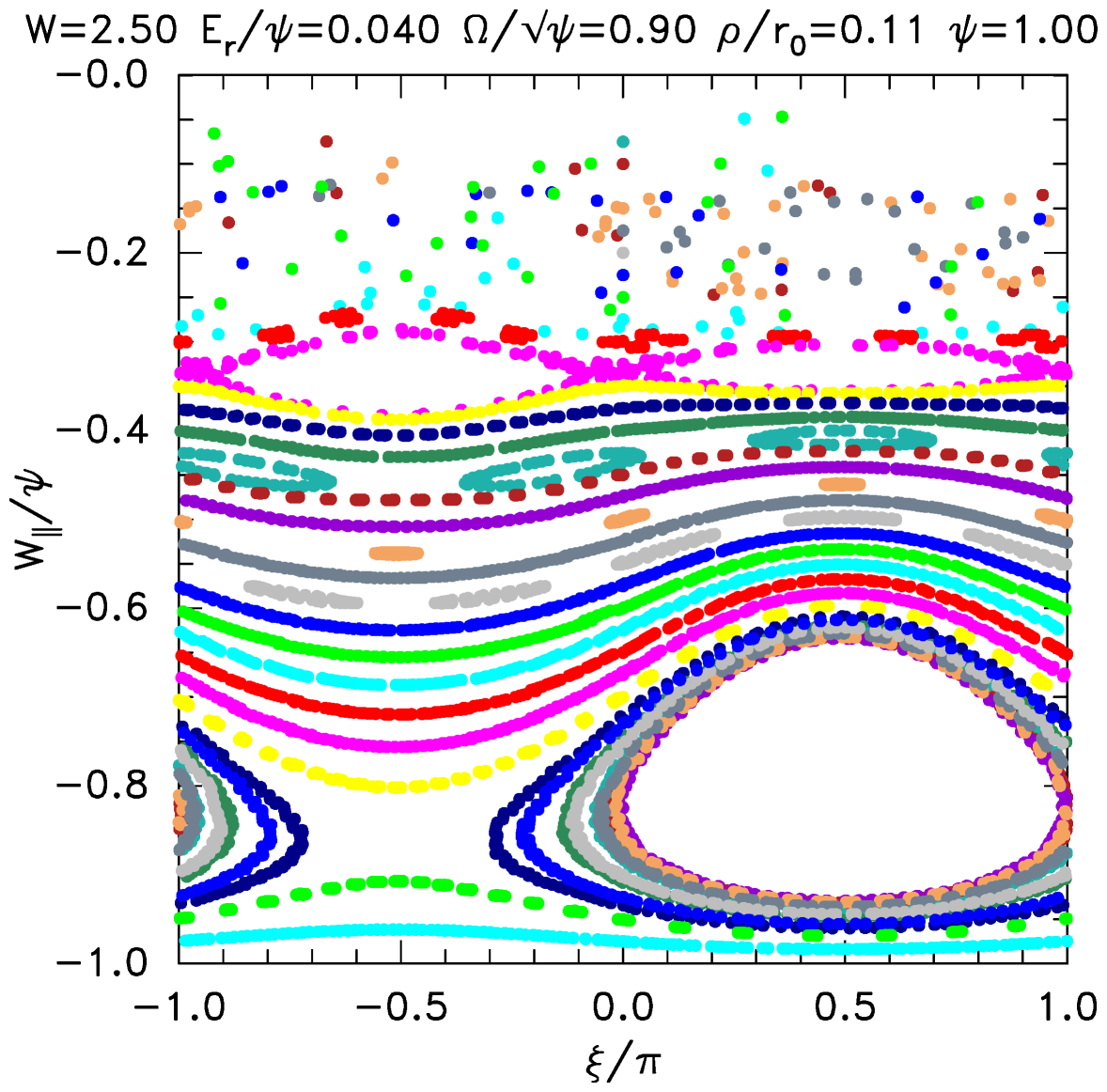}\hskip-2em (b)\phantom{M}}
  ;\end{tikzpicture}
\else
\includegraphics[width=\hsize]{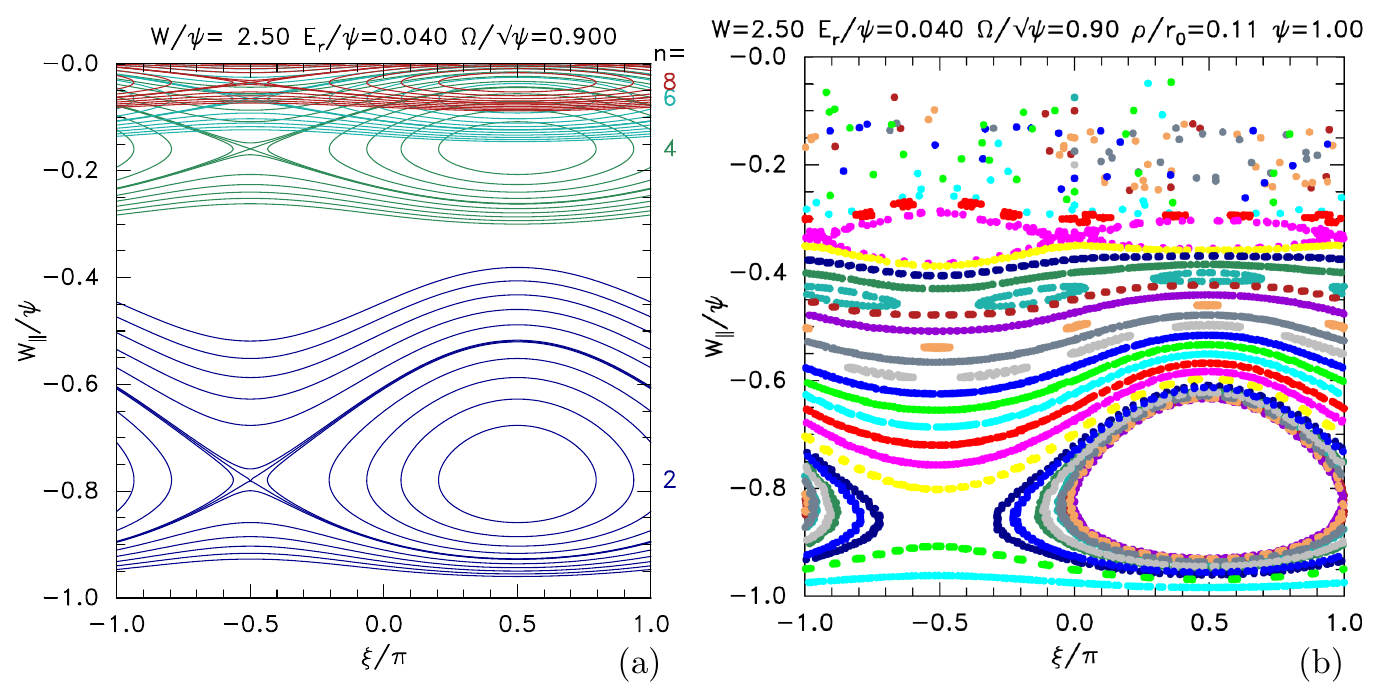}
\fi
  \caption{(a) Analytic energy trajectories for different harmonics
    ($n$), and fixed magnetic field strength and (b) Poincar\'e plot of
    the corresponding numerically integrated orbit, for a stronger
    transverse electric field.}
  \label{fig:traj940}
\end{figure}

Fig.\ \ref{fig:traj940}, by comparison, shows what happens if the
amplitude of the perturbing transverse field is increased by a factor
of 8, other parameters unchanged. The $n=4$ and $6$ islands now
overlap strongly, and the entire region $W_\parallel/\psi\gtrsim -0.3$
becomes stochastic. Below it, the Poincar\'e plots show orbits to be
permanently trapped. Small island chains with higher mode numbers in
phase $\xi$ become visible. For example the chain of 3 islands at
$W_\parallel/\psi\simeq-0.45$, or of two islands at
$W_\parallel/\psi\simeq-0.34$. These additional chains arise from
nonlinearity, and are not represented in the analytic linearized
approximation. Still, the overall extent of the $n=2$ island is quite
well captured by the analytics, which predict that it should remain
intact, as it does.
\begin{figure}
\ifextgen
  \tikzsetnextfilename{Figure8}
  \begin{tikzpicture}\node (image)
  {\noindent
  \includegraphics[width=0.491\hsize]{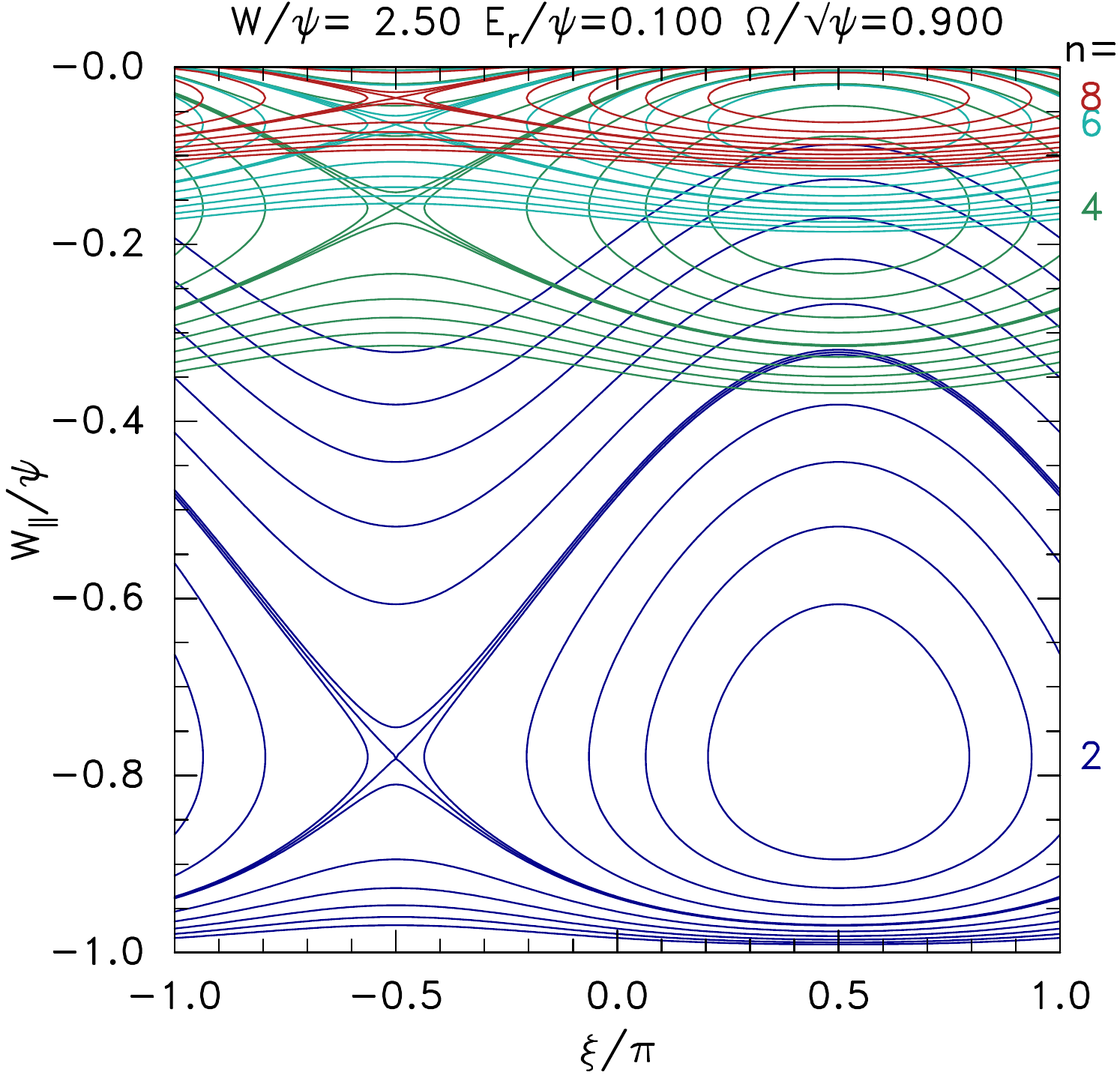}\hskip-2em (a)\hskip1em 
  \includegraphics[width=0.488\hsize]{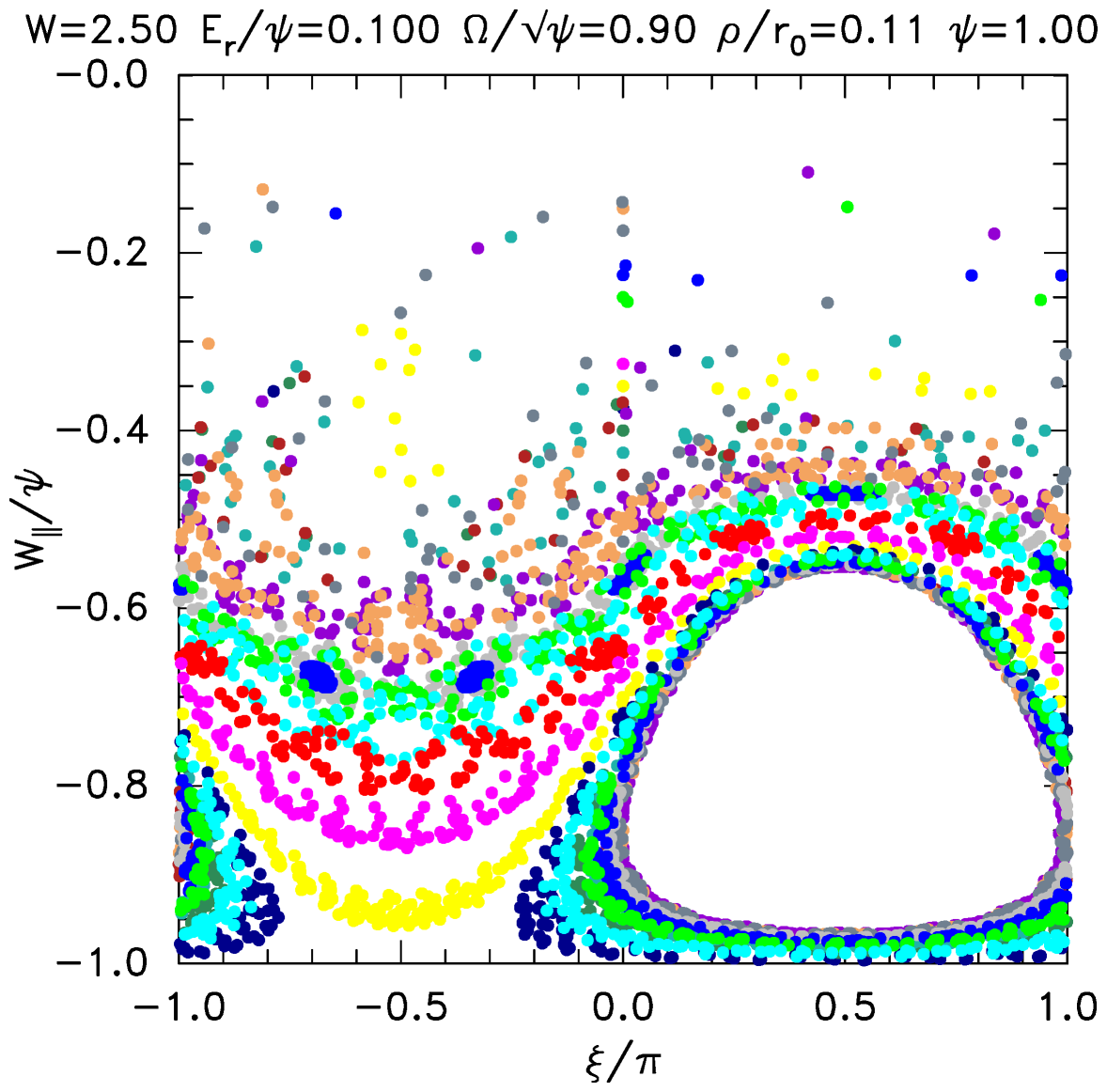}\hskip-2em (b)\phantom{M}}
  ;\end{tikzpicture}
\else
\includegraphics[width=\hsize]{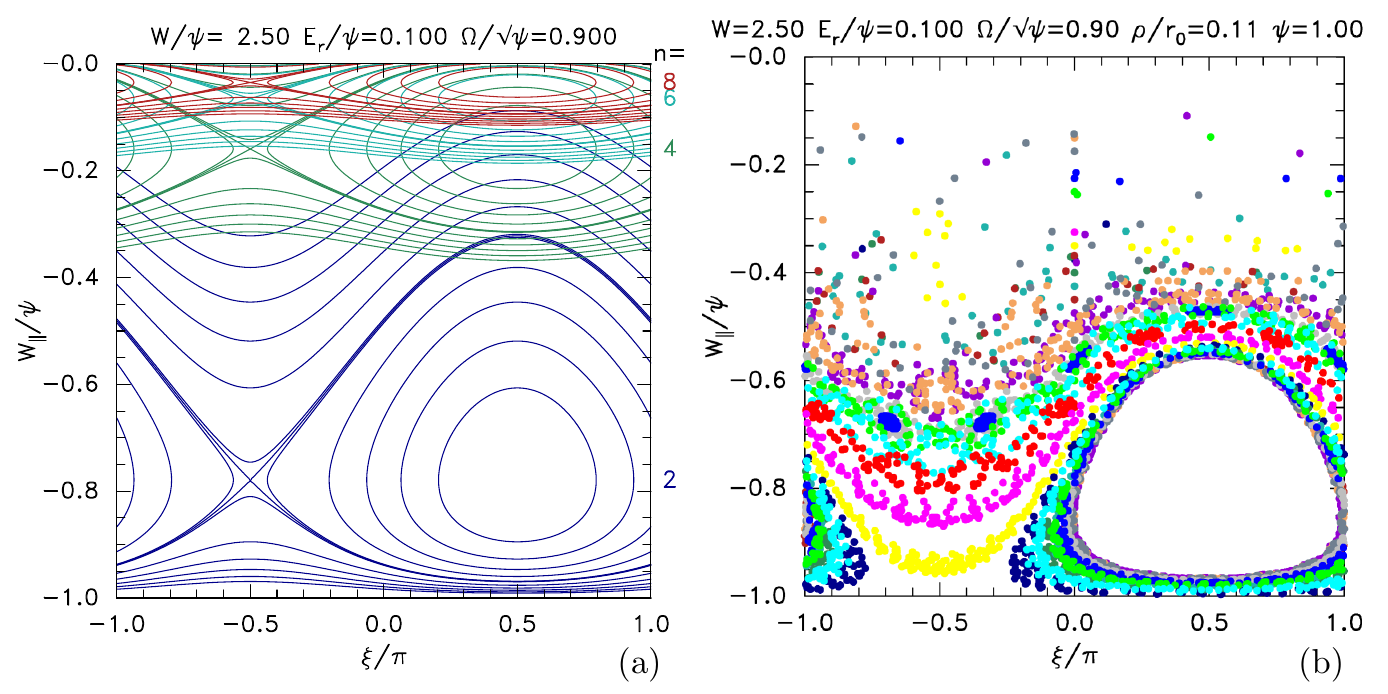}
\fi
  \caption{(a) Analytic energy trajectories for different harmonics
    ($n$), and fixed magnetic field strength and (b) Poincar\'e plot of
    the corresponding numerically integrated orbit, for an extremely strong
    transverse electric field.}
  \label{fig:traj9100}
\end{figure}
If the perturbing $E_r/\psi$ is increased to 0.1,  then
overlap and stochasticization of even the $n=2$ island begins, as
illustrated by Fig.\ \ref{fig:traj9100}. Soon beyond it, by $E_r/\psi=0.13$,
essentially the whole of the phase space becomes stochastic.

\begin{figure}
\ifextgen
  \tikzsetnextfilename{Figure9}
  \begin{tikzpicture}\node[anchor=south west] (image) 
  {\noindent
  \includegraphics[width=0.491\hsize]{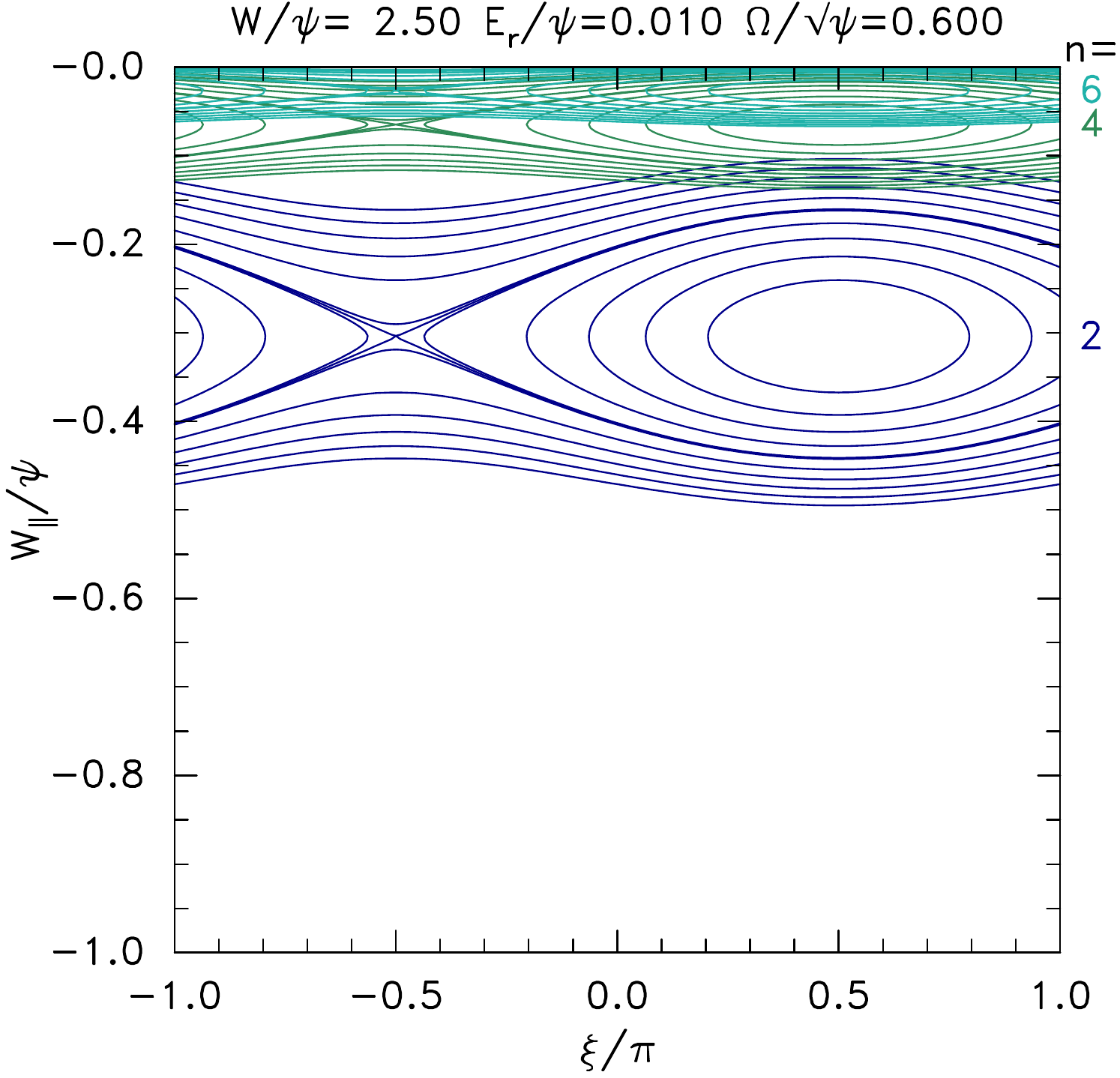}\hskip-2em (a)\hskip1em 
  \includegraphics[width=0.488\hsize]{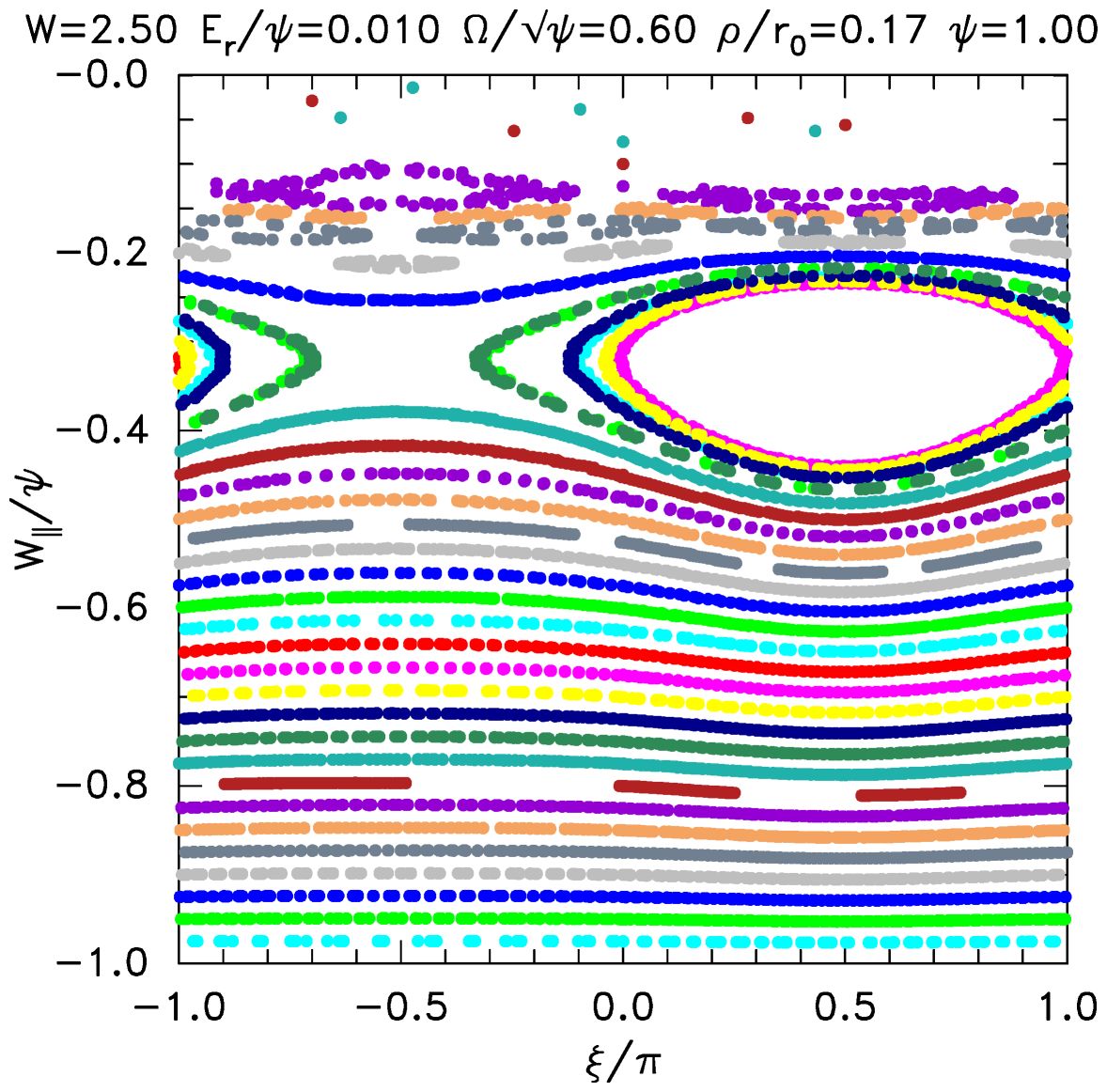}\hskip-2em (b)\phantom{M}};
 \node[anchor=north west] (image)
  {\includegraphics[width=0.491\hsize]{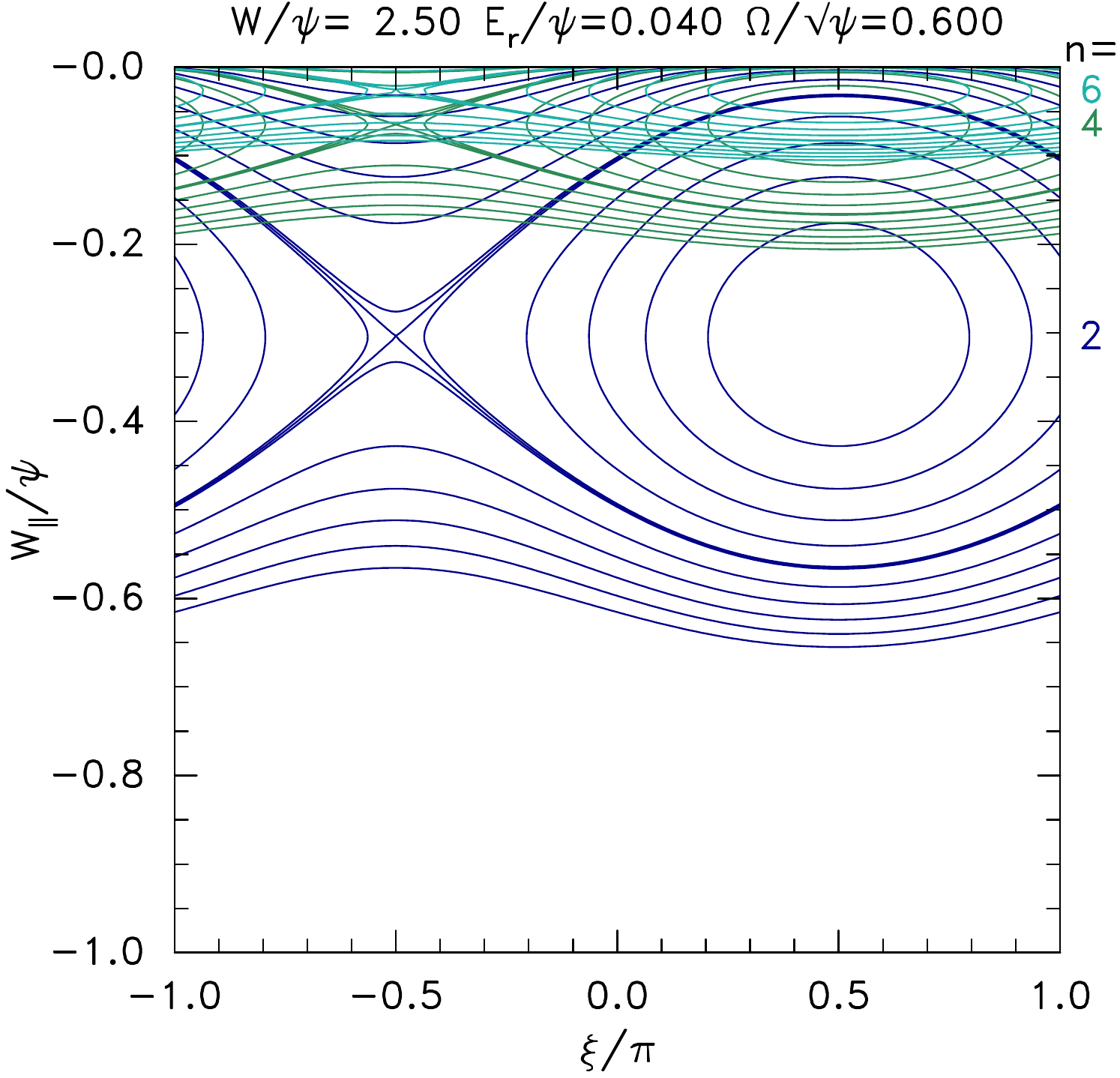}\hskip-2em (c)\hskip1em 
  \includegraphics[width=0.488\hsize]{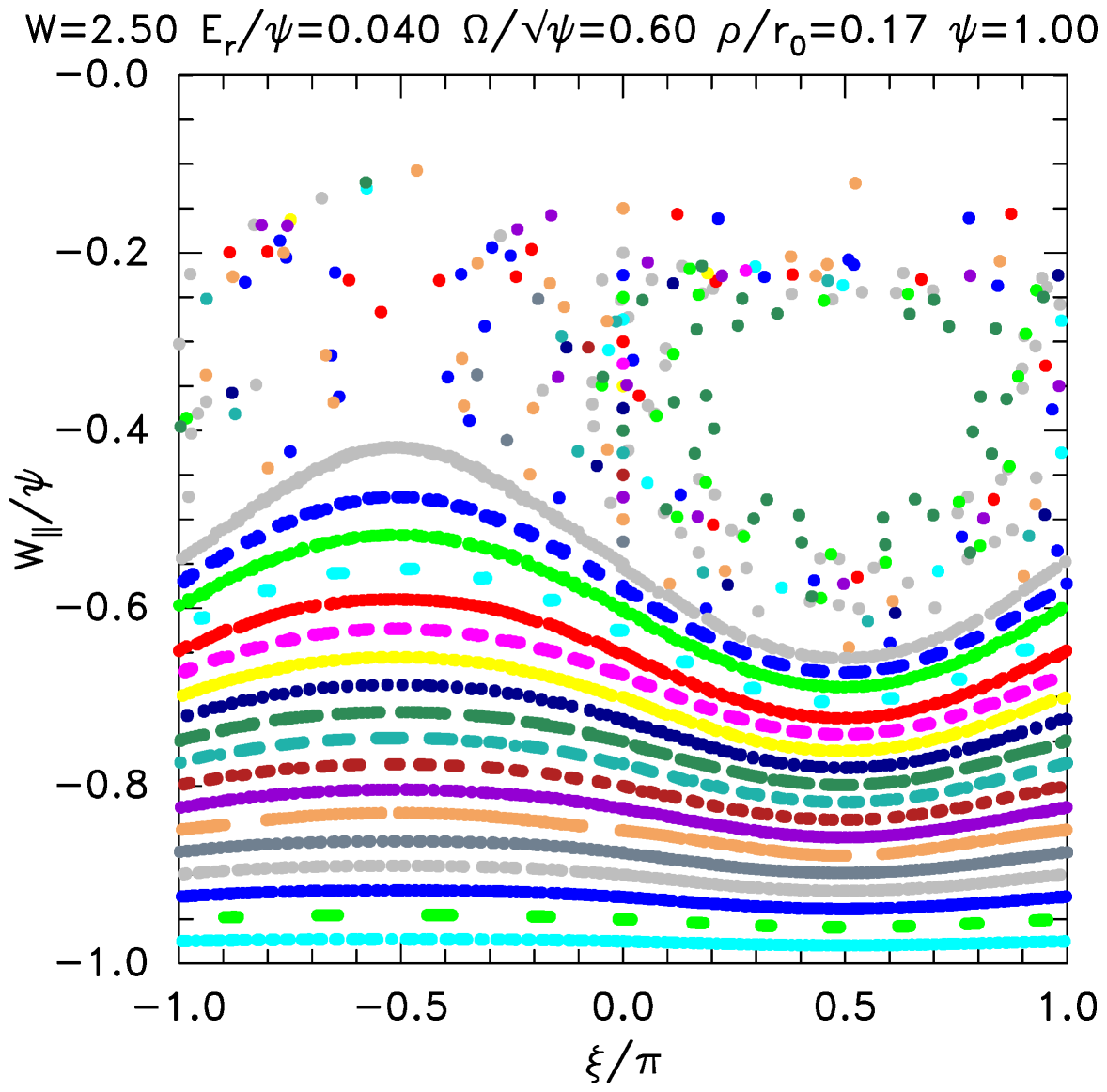}\hskip-2em (d)\phantom{M}}
  ;\end{tikzpicture}
\else
\includegraphics[width=\hsize]{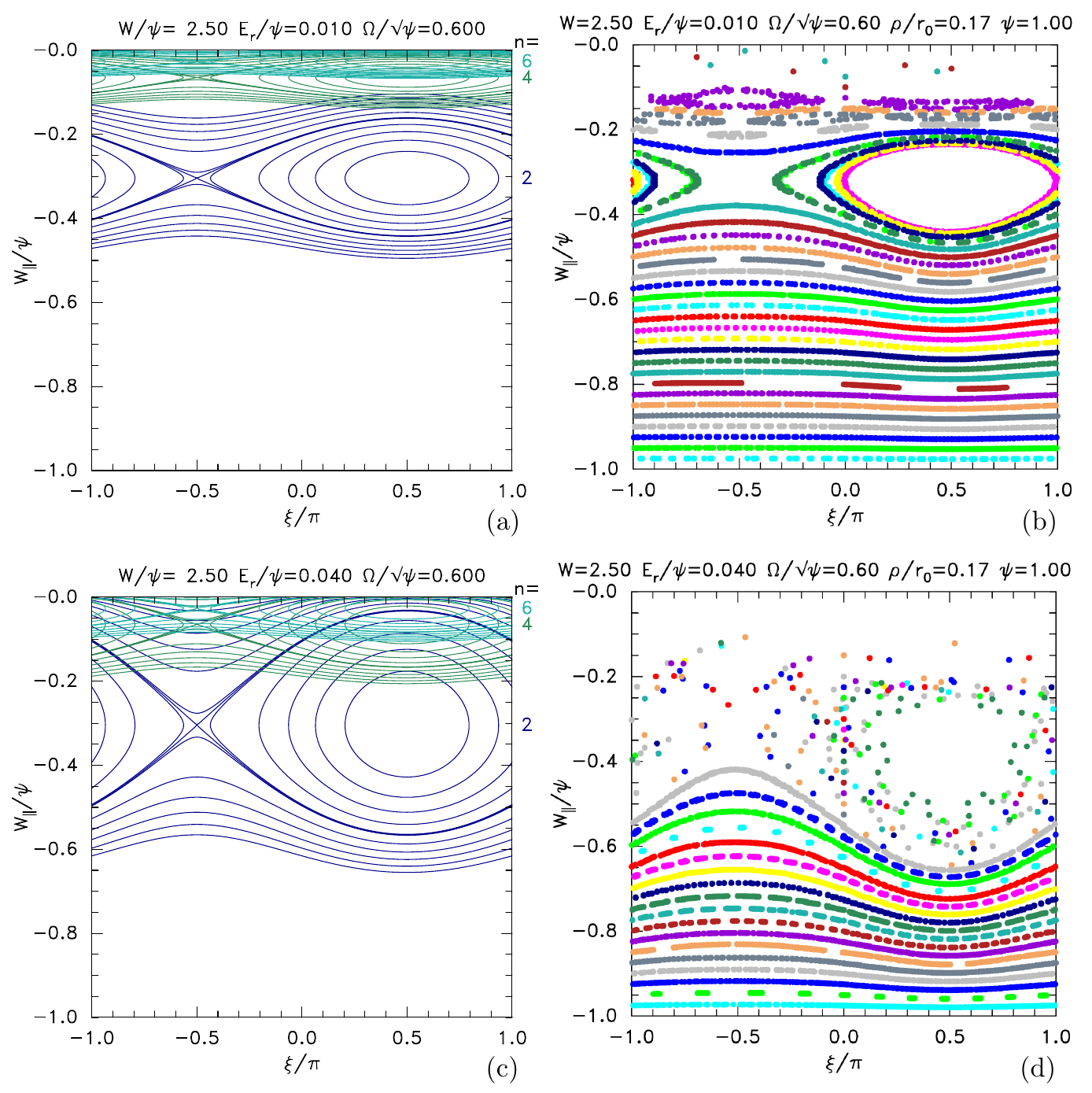}
\fi
  \caption{(a),(c) Analytic energy trajectories fixed magnetic field
    strength and (b),(d) Poincar\'e plots of the corresponding
    numerically integrated orbits, for a lower magnetic field, and two
    perturbation amplitudes.}
  \label{fig:traj6100}
\end{figure}
Fig.\ \ref{fig:traj6100} shows what happens for a lower magnetic
field, $\Omega/\sqrt{\psi}=0.6$. In this case, a field $E_r/\psi=0.04$
is sufficient to make the $n=2$ island stochastic, but when that happens,
there still remain some permanently trapped orbits at energies
sufficiently below the resonance value ($\sim0.6^2$).

\begin{figure}
\ifextgen
  \tikzsetnextfilename{Figure10}
  \begin{tikzpicture}\node[anchor=south west] (image) 
  {\noindent
  \includegraphics[width=0.491\hsize]{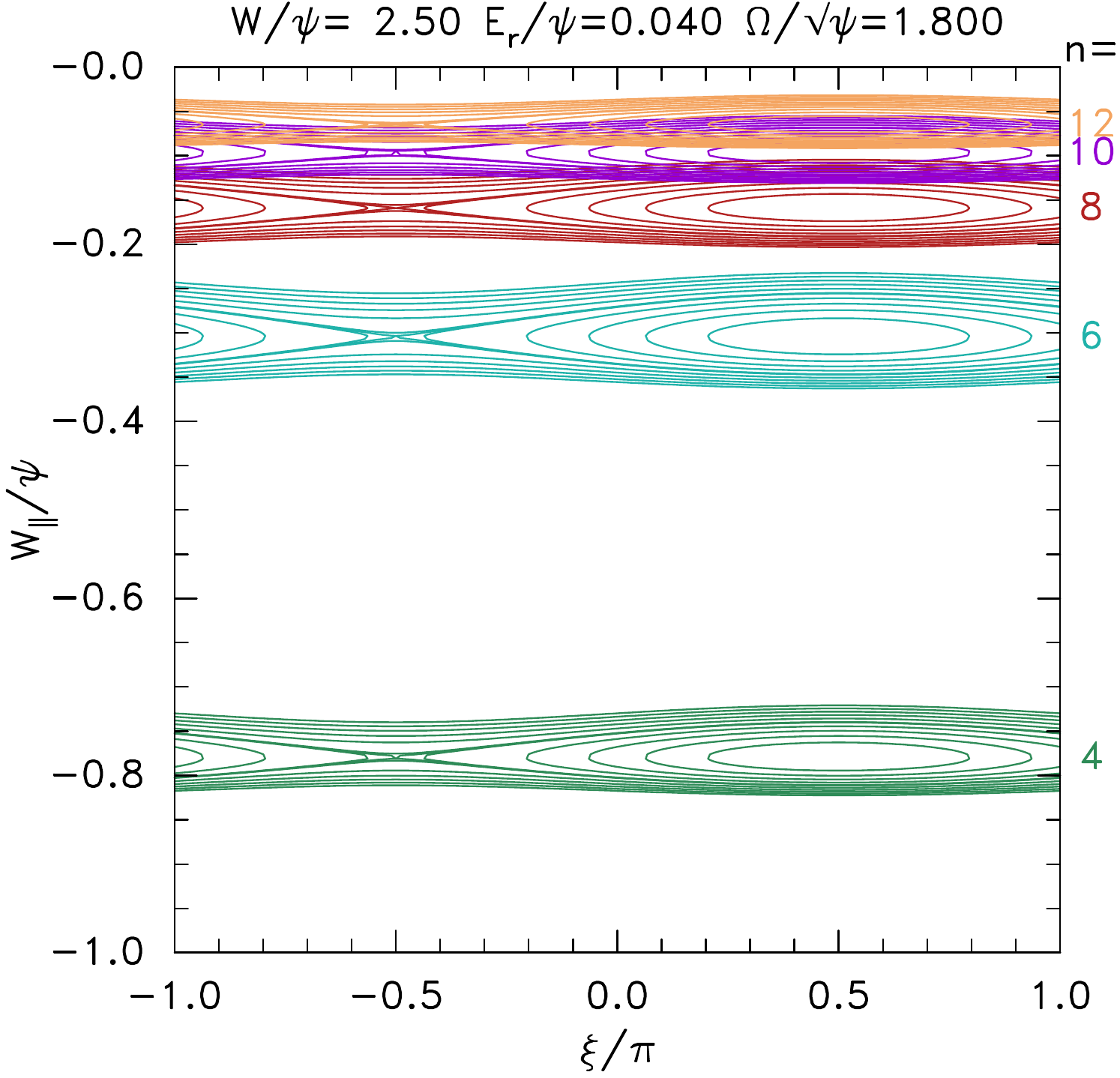}\hskip-2em (a)\hskip1em 
  \includegraphics[width=0.488\hsize]{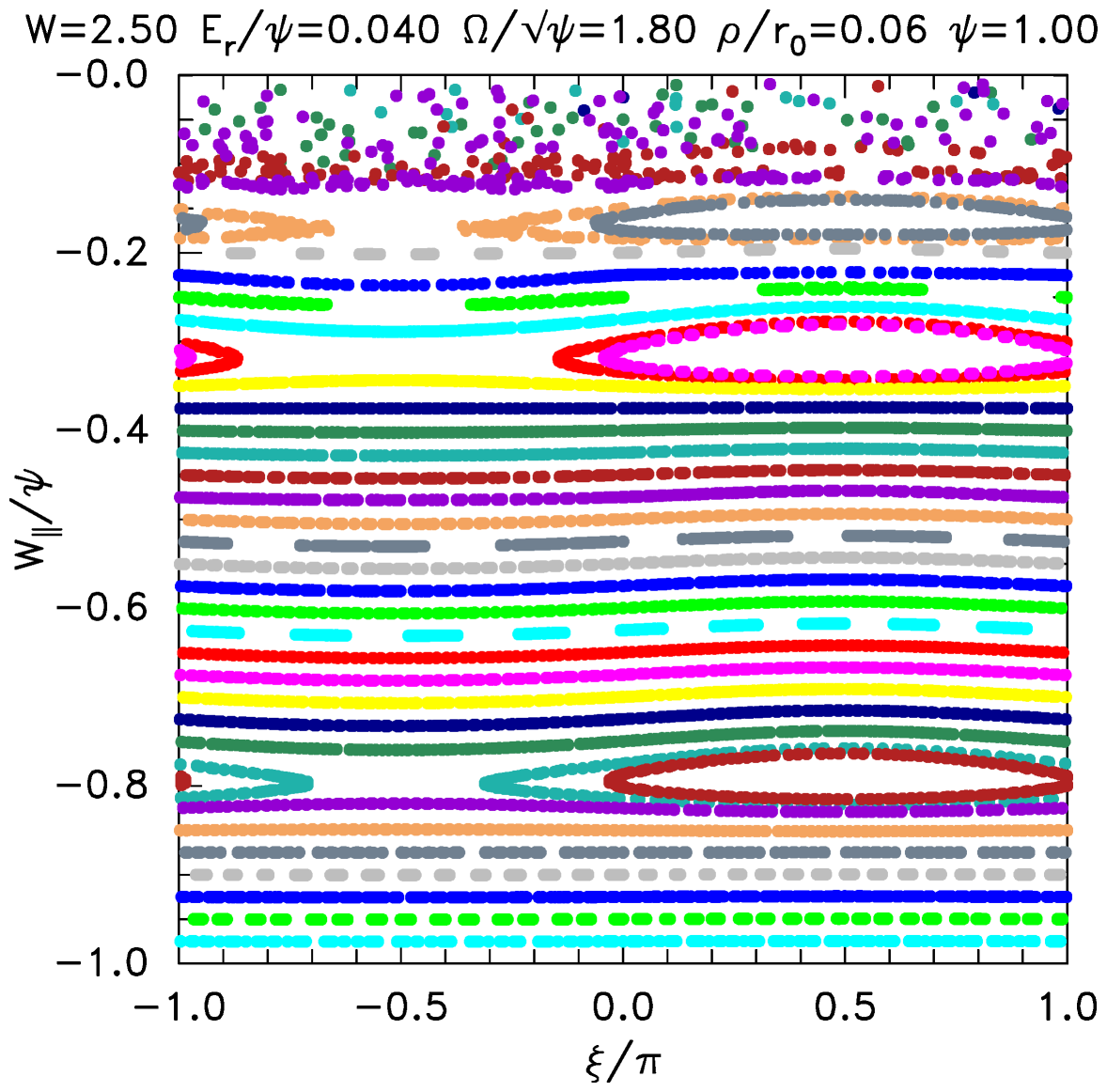}\hskip-2em (b)};
  \node[anchor=north west] (image)
  {\includegraphics[width=0.491\hsize]{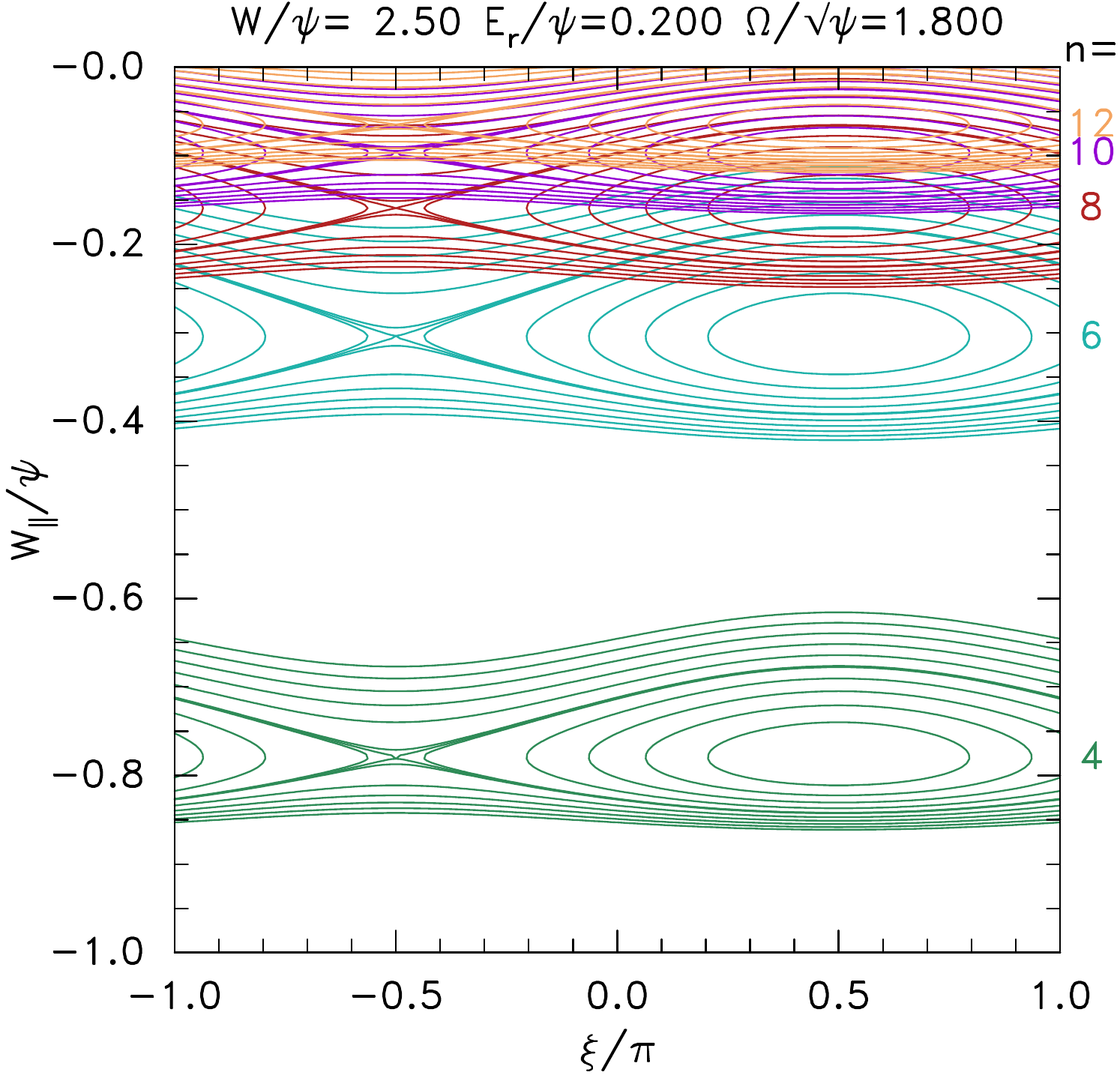}\hskip-2em (a)\hskip1em 
   \includegraphics[width=0.488\hsize]{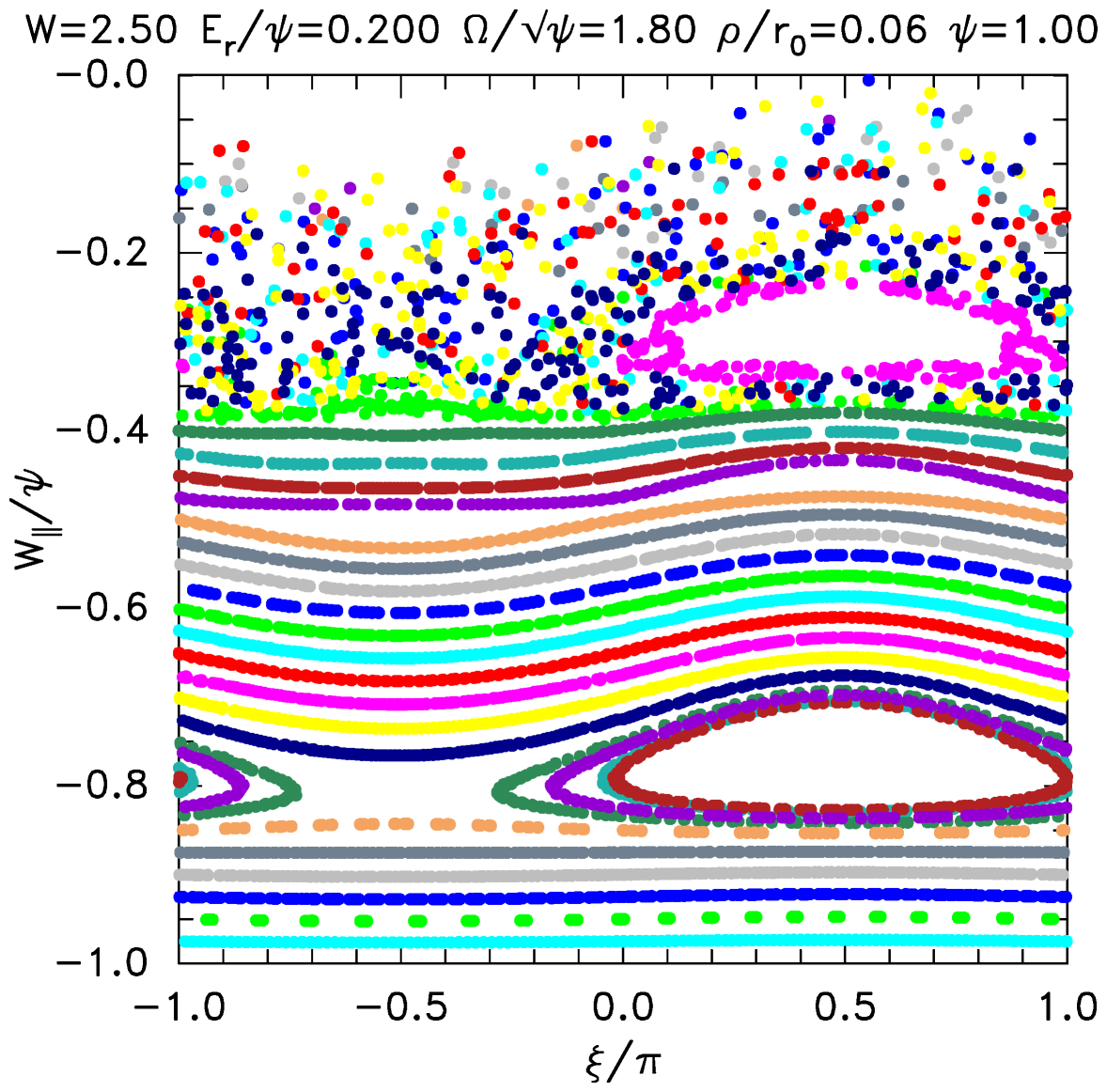}\hskip-2em (b)\phantom{M}}
  ;\end{tikzpicture}  
\else
\includegraphics[width=\hsize]{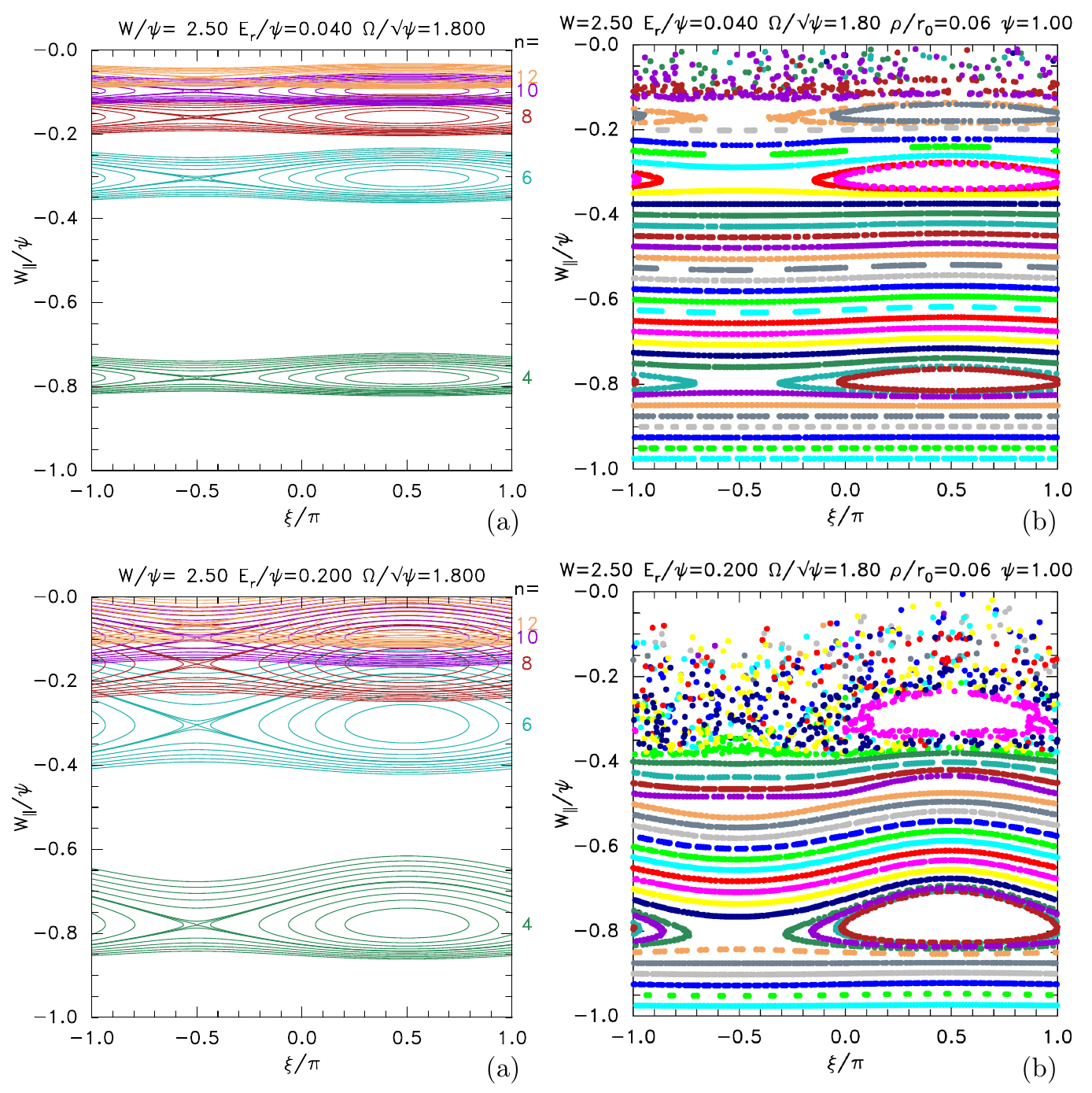}
\fi
  \caption{Analytic energy trajectories (a,c), and corresponding
    Poincar\'e plots (b,d), for $\Omega/\sqrt{\psi}=1.8$, and
    $E_r/\psi=0.04$ (a,b) or 0.2 (c,d).}
  \label{fig:traj1840}
\end{figure}

In contrast, as shown in Fig.\ \ref{fig:traj1840},
increasing the magnetic field to
$\Omega/\sqrt{\psi}=1.8$, removes the $n=2$ resonance; and because the
higher resonances are weaker, the orbits can sustain higher $E_r/\psi$
before becoming stochastic. This stabilizing effect is enhanced by the resulting
reduction in gyroradius $\rho$.

\section{Island widths,  overlap, and trapped phase space}

In the previous section we have shown that the island overlap
criterion successfully predicts which trajectories are stochastic (and
hence become untrapped) and which are permanently trapped. We
therefore rely on this success and formulate an analytic condition for
particles at different locations in phase space to be permanently
trapped. We will take those parallel energies $W_\parallel$ to be
trapped which lie \emph{below the bottom of the lowest overlapped
  island} and all others to be subject to detrapping. This criterion
describes within typically 10\% in $W_\parallel$ what has been
observed in the example cases we have shown.

The function $F_n$, for fixed $w$ and $\wr $, is stationary
at resonance ($\sqrt{\wp}=\sqrt{\wr }$), and its
derivative in the vicinity of the resonance can be taken from eq.\
(\ref{eq:combtraj1}) as
\begin{align}
  \label{eq:Fnprime}
  {\partial F_n\over \partial \sqrt{\wp}} 
  &= 2\sqrt{\wp}{\partial F_n\over \partial \wp}\\
  &=
   {n(\sqrt{\wp}-\sqrt{\wr })\over\sqrt{2} \sqrt{w+\wp}}
   \left[{\sqrt{\wp}\,n/2\over (1-\sqrt{\wp})^m} + 
     {\pi\over8}\right].\nonumber
\end{align}
Consequently the width of the island
separatrix, which occurs at $\xi=\pi/2$, is determined by the
$\sqrt{\wp}$ value for which
$F_n(\wp)-F_n(\wp=\wr )\simeq
{1\over2}(\sqrt{\wp}-\sqrt{\wr })^2
{\partial^2 F_n\over \partial \sqrt{\wp}^2}  $
is equal to $E_{r0}/\psi$.
Therefore, regarding the second derivative as constant (adopting just the
second-order term in a Taylor expansion of $F_n$) we can express the island
(half-) width as
\begin{align}
  \label{eq:halfwidth}
 \delta_n
&\equiv\sqrt{\wp}-\sqrt{\wr } \\
&\simeq
\left[  {E_{r0}\over
    \psi}{2\sqrt{2}\over n}{\sqrt{w+\wr }}\right]^{1/2}
 \left[{\sqrt{\wr }\,n/2\over (1-\sqrt{\wr })^m} + 
     {\pi\over8}\right]^{-1/2}.\nonumber
\end{align}
We write $w+\wr =w_\perp$
($\sqrt{2w_\perp}=v_\perp/\sqrt\psi$) and recognize that together the
parameters $n$, $E_{ro}v_\perp/\psi^{3/2}$, and $\wr$ determine
$\delta_n$ as follows.

\noindent
\textbf{Analytic Algorithm\quad}
Eq.\ (\ref{eq:resen}) $\wrn=[(n^2/2b^2)^{1/4}+1-2^{1/4} ]^{-4}$
enables us to find the energy of the upper and lower island limits of
island $n$ as
\begin{align}
  \label{eq:uplowlim}
  &\sqrt{\wrn}\pm\delta_n= \\
&\sqrt{\wrn} \pm\left[  {E_{r0}v_\perp\over
    \psi\sqrt{\psi}}{2\over n}\right]^{1/2}
 \left[{\sqrt{\wrn }\,n/2\over (1-\sqrt{\wrn })^m} + 
     {\pi\over8}\right]^{-1/2}.\nonumber 
\end{align}
It is in this equation that one must use the adjustment of $m$ of eq.\
(\ref{eq:adhocadj}) for high harmonics. Overlap occurs between the $n$
and $n+2$ harmonic islands when
$\sqrt{w_{\parallel R n}}-\delta_n<\sqrt{w_{\parallel R
    n+2}}+\delta_{n+2}$.
Beginning at the lowest value of $n$ for which a resonance exists
(requiring $\wrn<1$) determine from evaluation of
$\sqrt{w_{\parallel R n}}-\delta_n$ and
$\sqrt{w_{\parallel R n+2}}+\delta_{n+2}$ whether it overlaps with the
$n+2$ island. If so, then it is the \emph{lowest energy overlapped
  island}; if not, increment $n$ by 2 and repeat until overlap is
found. The resulting $n$ is the harmonic whose island's lower energy
limit is sought, which is
\begin{equation}\label{eq:wpt1}
W_{\parallel t}/\psi=-w_{\parallel t}=-(\sqrt{\wrn}+\delta_n)^2.
\end{equation}
Energies below this approximate bound are predicted trapped, energies
above have stochastic orbits and are detrapped.

\begin{figure}
\ifextgen
  \tikzsetnextfilename{Figure11}
  \begin{tikzpicture}\node (image)
  {\noindent
  \includegraphics[width=0.52\hsize]{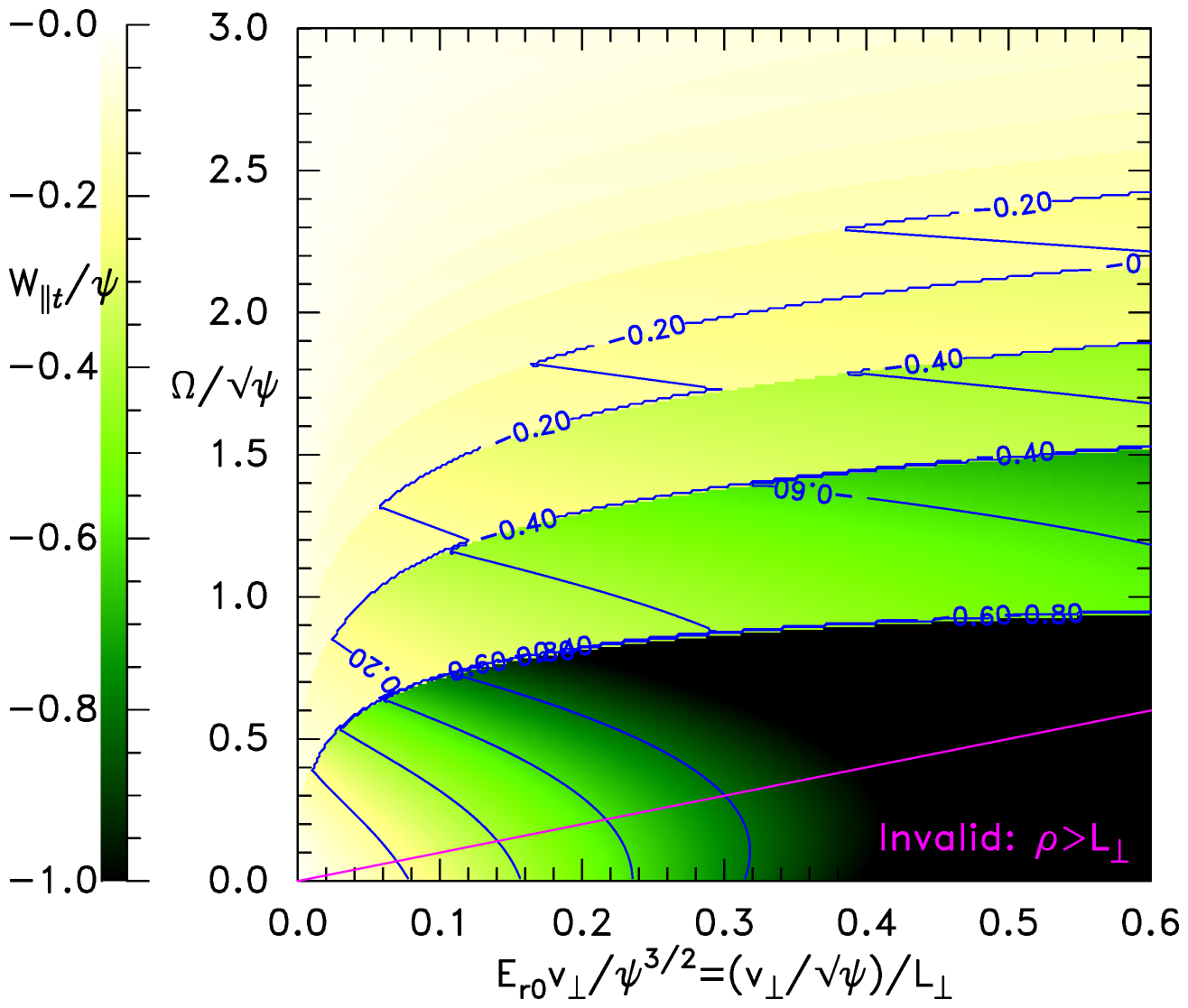}\hskip-2em(a)\hskip1em
  \includegraphics[width=0.46\hsize]{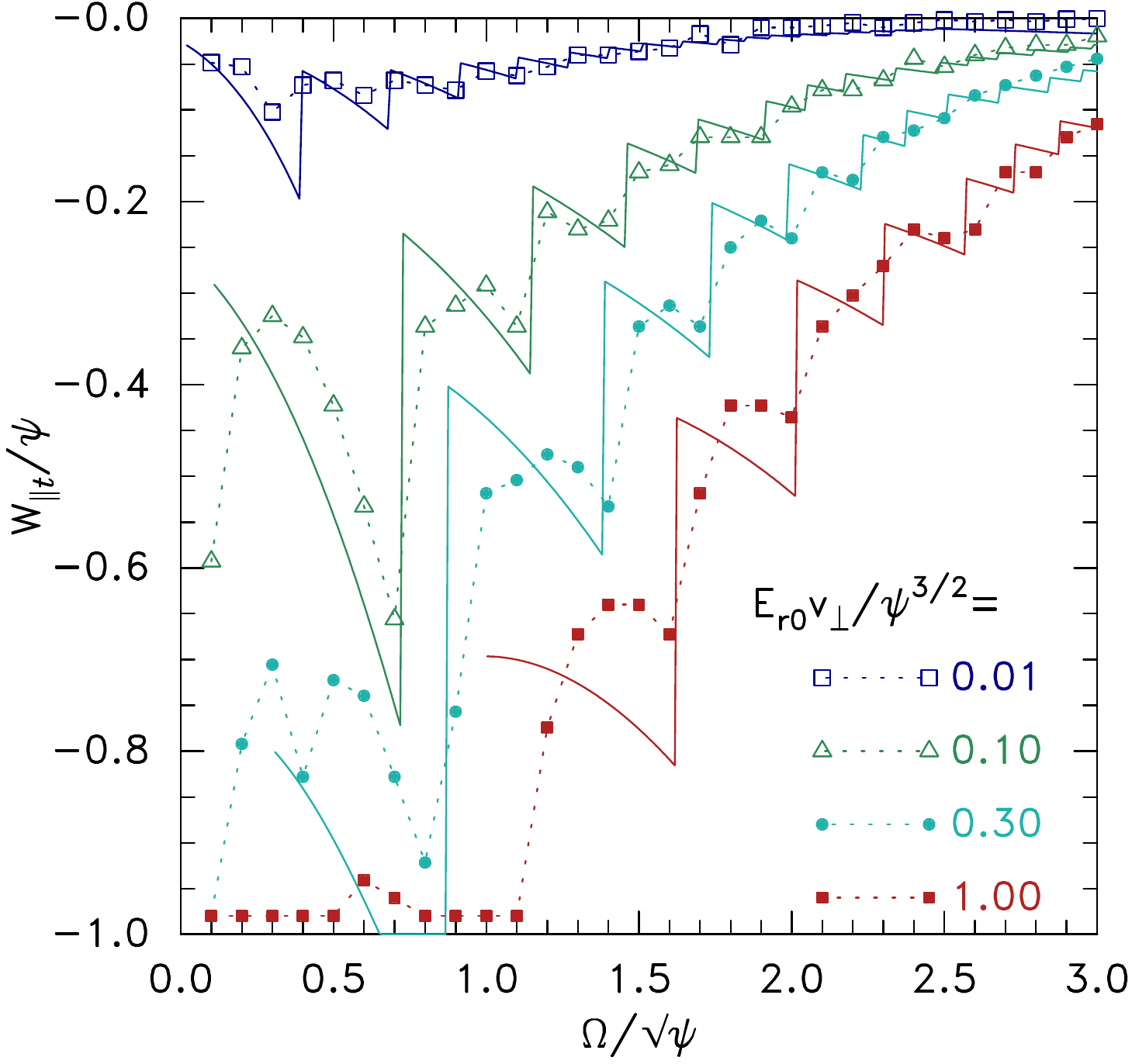}\hskip-2em(b)\phantom{M}}
  ;\end{tikzpicture}
\else
\includegraphics[width=\hsize]{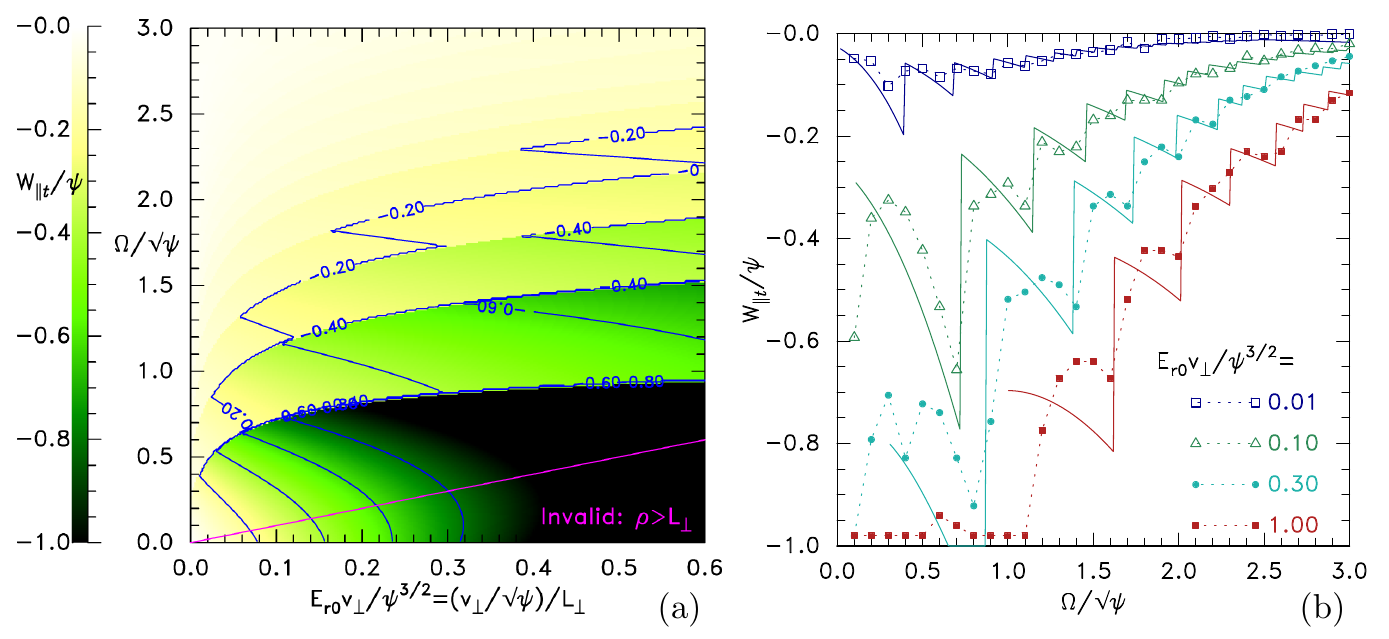}
\fi
  \caption{(a) Contours of the energy boundary $W_{\parallel t}$
    between trapped and detrapped orbits as a function of perturbation
    strength $E_{r0}v_\perp/\psi^{3/2}$ and magnetic field
    $\Omega/\sqrt{\psi}$. (b) The energy boundary $W_{\parallel t}$
    between trapped and detrapped orbits from eq.\ (\ref{eq:wpt1}) compared with the lowest
    detrapped orbits found from numerical orbit integration.}
  \label{fig:trapdomain}
\end{figure}

Figure \ref{fig:trapdomain}(a) shows the universal contours that result.
Where $W_{\parallel t}/\psi$ is close to zero (light regions), very
few orbits are detrapped; while where $W_{\parallel t}/\psi$ is close
to -1 (dark regions) almost all orbits are detrapped. Discontinuities
in $W_{\parallel t}$ occur where $n$ changes: it starts at 2 at the
bottom (right, below $\Omega/\sqrt{\psi}\simeq 1$) and increments
through 4,6,\dots\ as one moves to larger $\Omega/\sqrt{\psi}$. To
avoid almost complete detrapping for $\Omega/\sqrt{\psi}\lesssim 1$,
extremely weak perturbation is required. In contrast, for
$\Omega/\sqrt{\psi}\gtrsim 2$ there is a substantial region of
permanently trapped orbits even up to the largest perturbation
strength shown.

In Fig.\ \ref{fig:trapdomain}(b) are shown vertical profiles through
the contours at four values of the perturbation strength, giving
$W_{\parallel t}$ as a function of $b$. These lines are each
accompanied by points, each of which comes from full numerical orbit
integration. A point gives the lowest starting energy that escapes
during the first 200 bounces (which might take as many as a million
time-steps). We observe that there is very good agreement (even in
respect of the discontinuities) between the points and the lines. 

A more approximate form of the island widths can be obtained by using
$m=n/2$, substituting the more approximate frequency fit
$\wr =4\Omega^2/n^2\psi=4b^2/n^2$ so that
$\sqrt{w_{Rn}}-\sqrt{w_{Rn+2}}\simeq 4b/n^2$, and approximating
$(1-\sqrt{\wr })^{n/2} =(1-2b/n)^{n/2}\simeq {\rm e}^{-b}$. Then we
find $\delta_n$ is approximately proportional to $1/\sqrt{n}$ and can
be written
\begin{equation}
  \label{eq:halfomega}
 \delta_n\simeq\left[  {E_{r0}\over
    \psi}{2\sqrt{2w_\perp}\over n}\right]^{1/2}
 \left[
b{\rm e}^b
     +{\pi\over8}\right]^{-1/2}
\end{equation}
With reference to this approximation, the behavior can readily be
understood as follows. Island overlap
($2\delta_n\gtrsim \sqrt{w_{Rn}}-\sqrt{w_{Rn+2}}$) leading to
stochastic trajectories occurs if $\delta_n$ is too large, that is if
$E_{r0}v_\perp/\psi$ is too large provided $b$ $(=\Omega/\sqrt\psi)$ is not
large; or else if $n$ is too large, making
$\sqrt{w_{Rn}}-\sqrt{w_{Rn+2}}$ too small. The last of these cases (high $n$
at modest $E_{r0}$ and $b$) predicts that there is in principle
\emph{always} a stochastic region at very small $\wp$, where the
bounce frequency is correspondingly small and the resonant bounce
harmonic number large, regardless of the exact $E_{r0}$ and $b$
values. Consequently, a steady electron hole of limited transverse
extent will always have a stochastic transition between trapped and
passing orbits that in practice smooths out any steep $f$-gradients at
the separatrix. Our numerical orbit integration confirms this
prediction.

When $b$ ($=\Omega/\sqrt\psi$) is large, the term ${\rm e}^b$ makes
$\delta_n$ small, regardless of $E_{r0}$, and suppresses overlap. This
effect can be considered to arise because when the gyro-period is
small compared with the central transit time
($\tau_t\propto 1/\sqrt\psi$, the duration of the impulse), the
Fourier transform of a single impulse has become exponentially small
at the cyclotron frequency. The suppression applies at essentially all
$\wp$ up to 1, because the impulse width is a rather weak (slowly
increasing) function of $\wp$. Only the exponentially-large-$n$ orbits
at exponentially-small-$\wp$ will then be stochastic. And the region
of stochasticity is limited to very small $\wp$. High enough
magnetic field thus justifies the drift orbit treatment, and
eventually imposes no minimum $L_\perp$ requirement for a long-lived
hole to exist.

The opposite case $b\ll 1$ (weak magnetic field) preserves the assumed
localization in $r$ only if the transverse length scale remains
greater than the gyro-radius
$E_{r0}/\psi=1/L_\perp\lesssim
1/\rho=\Omega/v_\perp=b\sqrt\psi/v_\perp$
so $E_{r0}v_\perp/\psi^{3/2} \lesssim \Omega/\sqrt\psi$. The valid
region of Fig.\ \ref{fig:trapdomain}(a) is therefore above the
diagonal straight line $E_{r0}v_\perp/\psi^{3/2}= \Omega/\sqrt\psi$
drawn in purple. And in Fig.\ \ref{fig:trapdomain}(b), the lines are
drawn only in the valid region.  In the invalid region one can
expect the permanent trapping to be poor, and this is confirmed by the
points.

\begin{figure}
\ifextgen
  \tikzsetnextfilename{Figure12}
  \begin{tikzpicture}\node (image)
  {\noindent
  \includegraphics[width=0.48\hsize]{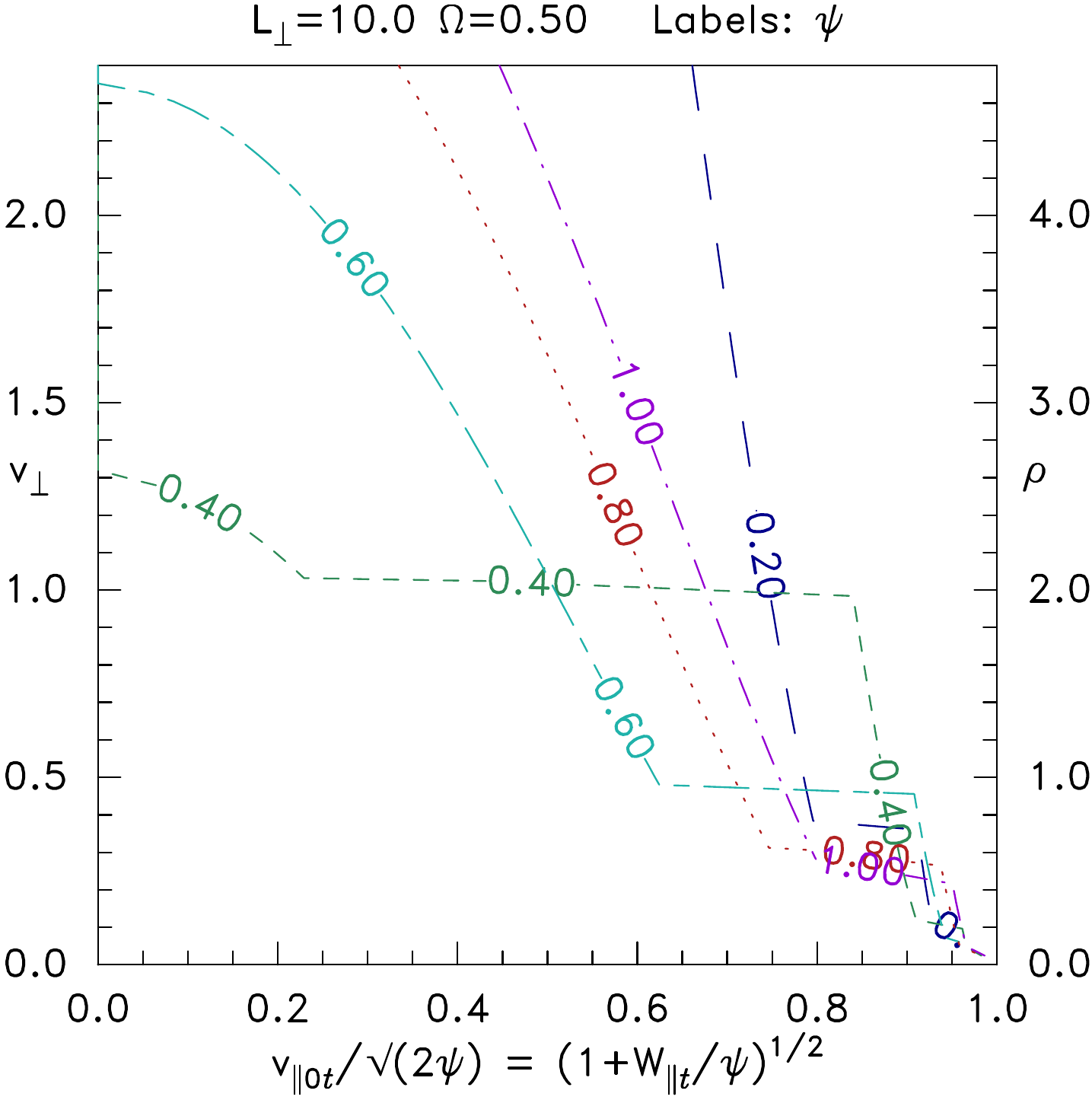}\hskip-2em (a)\hskip1em
  \includegraphics[width=0.48\hsize]{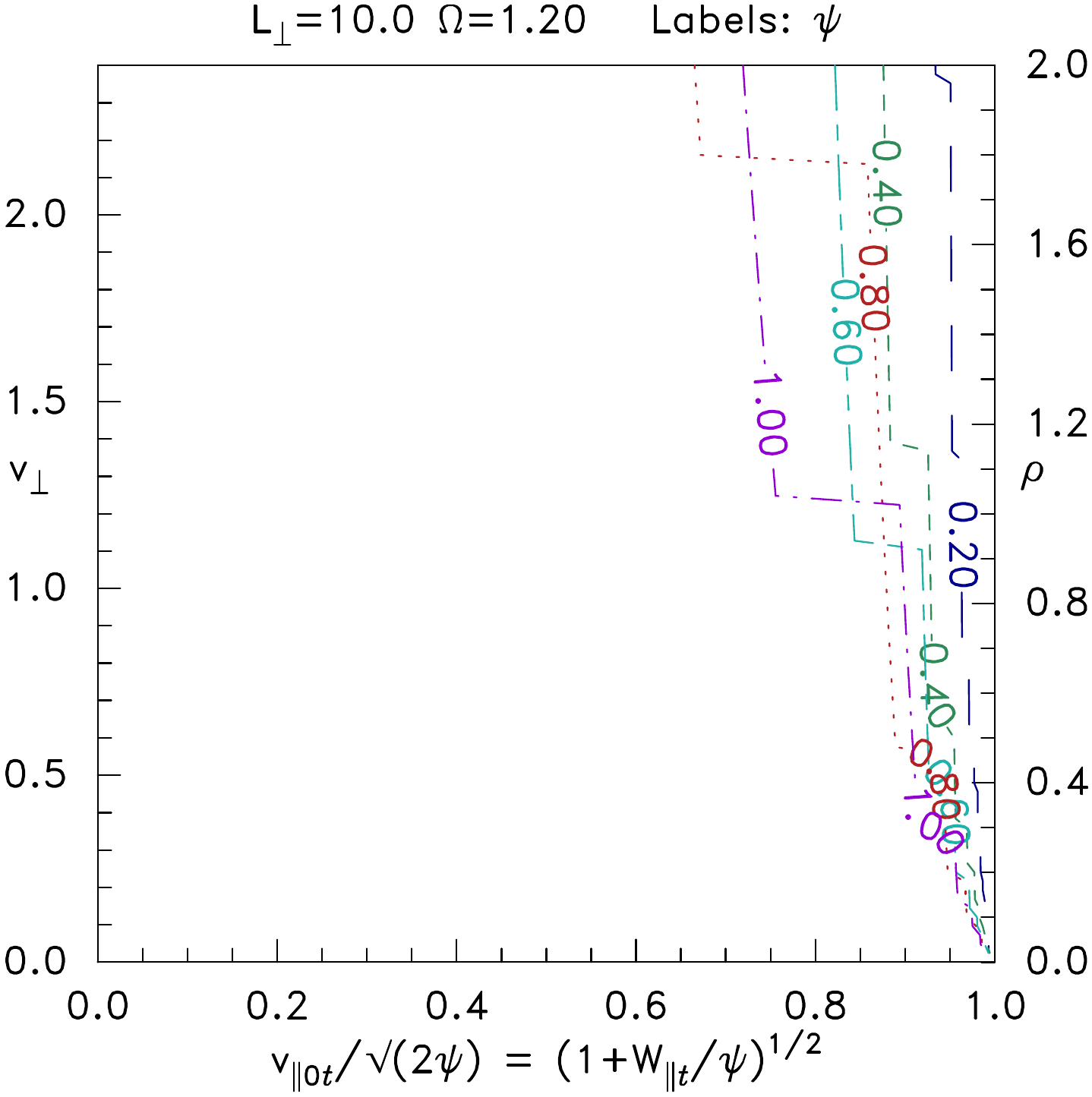}\hskip-2em (b)\phantom{M}}
  ;\end{tikzpicture}
\else
\includegraphics[width=\hsize]{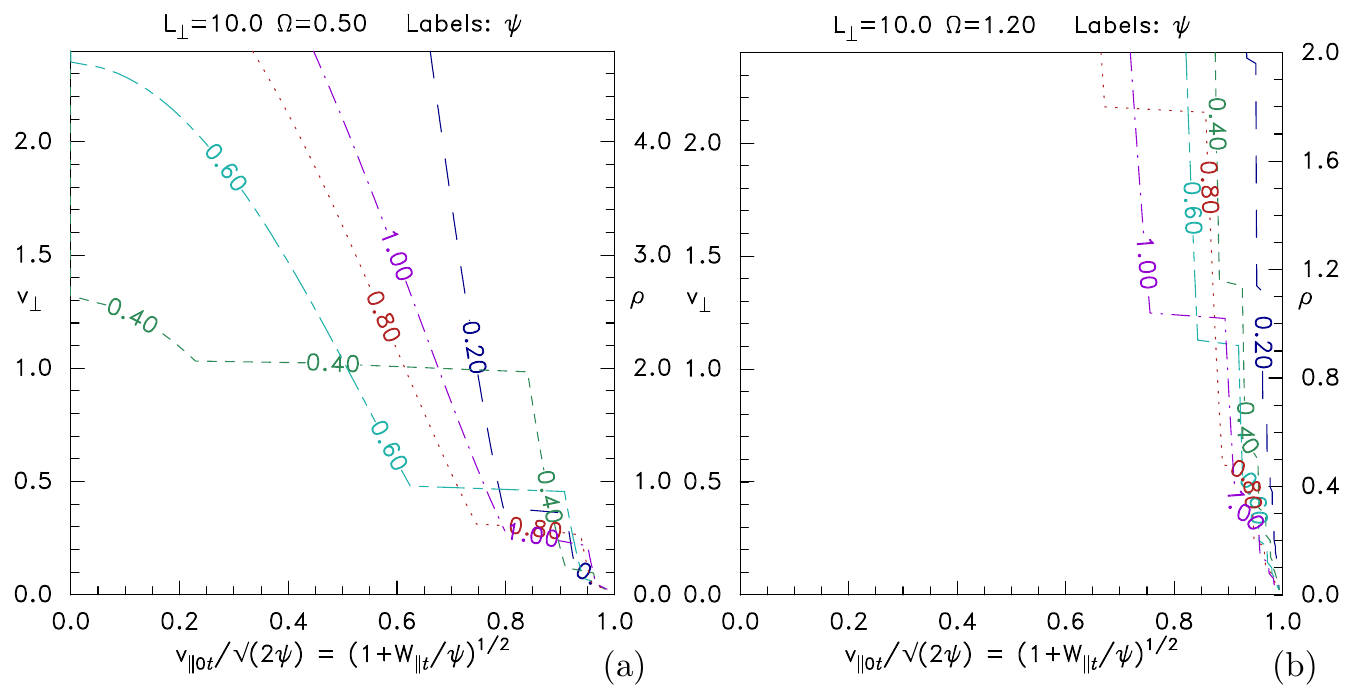}
\fi
  \caption{The boundary in velocity-space measured at $z=0$ between
    trapped and untrapped orbits at different $\psi$-values: (a) for
    low magnetic field $\Omega=0.5$, (b) for higher $\Omega=1.2$. }
  \label{fig:vspace}
\end{figure}

A perhaps more intuitive way to portray typical results is as in Fig.\
\ref{fig:vspace},
where are shown examples of boundaries between trapped and untrapped
orbits in velocity-space (based on the island overlap calculation).
The important regions of this domain extend to thermal velocities
($v_\perp\gtrsim 1$, not just $v_\perp \gtrsim \sqrt\psi$). We need
orbits to be permanently trapped for most of the range of possible
$v_\perp$ to allow the depression of $f(v_\parallel,v_\perp)$ to
contribute sufficient positive charge to sustain the hole. For smaller
$\psi$ the effective perturbation strength
$\propto v_\perp/ \sqrt{\psi}L_\perp$ becomes stronger for
given $L_\perp$, which makes orbits more easily detrapped. However,
the effects of varying resonance condition as $\psi$ changes are very
strong; so the boundaries do not behave monotonically with
$\psi$. When the $n=2$ resonance is avoided, as in Fig.\
\ref{fig:vspace}(b), the boundary lies at fairly high velocity near
$W_\parallel=0$. That leads us to expect qualitatively that a
distribution $f(v_{\parallel0})$ that is approxiately flat above
$v_{\parallel0t}$, in the stochastic region, can still sustain an
electron hole with these parameters.

In all cases, increasing $L_\perp$ and making the hole more oblate,
i.e.\ closer to one-dimensional reduces the detrapped phase-space
area. But unless $\Omega/\sqrt\psi\gtrsim 2$, holes of large
transverse dimension are unstable to transverse perturbations that
grow in a few hundred plasma periods and break up the holes into
shorter transverse lengths, causing them to collapse. So there is a
competition between the requirements of equilibrium and stability.

\section{Summary}

It has been shown that parallel energies of deeply trapped orbits in
axisymmetric electron holes have limited excursions in parallel
energy, provided the transverse electric field perturbation is weak
enough. There is a parallel energy threshold which is a function of
perturbation strength and magnetic field, above which the parallel
energy trajectory becomes stochastic, and is no longer limited in
extent, instead becoming detrapped. Such orbits cannot therefore
contribute to the electron deficit needed to sustain the hole. The
stochasticity arises when trajectory islands overlap, as has been
confirmed by numerical orbit integration. The parallel energy
threshold for detrapping has been quantitatively evaluated using the
\textbf{Analytic Algorithm} as a universal function of the hole
parameters. Magnetic fields strong enough that
$\Omega/\sqrt\psi\gtrsim 2$ allow a large fraction of the orbits with
negative parallel energy to be permanently trapped, even for quite
short transverse scale lengths. However, lower magnetic field
strengths $\Omega/\sqrt\psi\lesssim 1$ have most of their orbits
detrapped unless the transverse scale length is rather large. Although
fully self-consistent hole equilibria have not yet been calculated,
the present results appear to give an explanation based upon
equilibrium trapping constraints for the observation that holes with
lower magnetic field and lower peak potential generally must have
greater transverse extent than those with greater field or greater
potential. Future work will aim to use the quantitative results of
this trapped-phase-space calculation, illustrated in Figs.\
\ref{fig:trapdomain} and \ref{fig:vspace}, to explore when fully
self-consistent 2D holes can exist and what their forms are likely to
be.

\section*{Appendix: Mathematical Function Details}

The integrated expressions for $F_n$ are as follows
\begin{align}
  \label{eq:integrals}
  F_n&={n^2\over 2\sqrt{2}}g_m+{n\pi\over16\sqrt2} g_0,\\
\noalign{with}
  g_0&=\int {(\sqrt{\wp}-\sqrt{\wr })d\wp
       \over\sqrt{w+\wp}\sqrt{\wp}}; \qquad \\
  g_m&=\int  {(\sqrt{\wp}-\sqrt{\wr })d\wp
       \over\sqrt{w+\wp}(1-\sqrt{\wp})^m}.
\end{align}
The first function is easy: $g_0=2[\sqrt{w+\wp}-w\swr\ln(\sqrt{w+\wp}+\swp)]$.
To evaluate $g_m$, define the integrals
\begin{align}
  \label{eq:integrals2}
  I_m(a,x)=\int {dx\over \sqrt{a+x^2}(1-x)^m},\\ 
\qquad J_m(a,x)=\int {x dx\over \sqrt{a+x^2}(1-x)^m};
\end{align}
then, since $x^2=(x-1)x+x$, it is easy to show that
\begin{equation}
  \label{eq:gnfromj}
   g_m=2[(1-\swr)J_m(w,\swp)-J_{m-1}(w,\swp)].
\end{equation}
The $J_m$ and $I_m$ are related by 
\begin{equation}
  \label{eq:jnin}
   J_m(a,x)=\int {(x-1) +1 \over \sqrt{a+x^2}(1-x)^m}dx=I_m-I_{m-1}.
\end{equation}
Also $J_m$ can be integrated by parts as
\begin{align}
  \label{eq:jnparts}
  J_m&={\sqrt{a+x^2}\over (1-x)^m}
  -m\int {(1-x)^2-2(1-x)+1+a\over
       \sqrt{a+x^2}(1-x)^{m+1}}dx\nonumber\\
  &= {\sqrt{a+x^2}\over (1-x)^m} -mI_{m-1}+2mI_{m}-m(1+a)I_{m+1}.
\end{align}
Eliminating $J_m$ between (\ref{eq:jnin}) and (\ref{eq:jnparts}), and gathering
terms we obtain the following recursion relation:
\begin{equation}
  \label{eq:recursion}
  I_{m+1}=\left[{\sqrt{a+x^2}\over
      (1-x)^m}+(2m-1)I_m-(m-1)I_{m-1}\right]
  {1\over m(a+1)}.
\end{equation}
Given $J_0=\sqrt{a+x^2}$, and the initial values of the recursion:
$I_0=\ln(\sqrt{a+x^2}+x)$, and
$I_1=[\ln(\sqrt{a+1}\sqrt{a+x^2}+a+x)-\ln(1-x)]/\sqrt{a+x^2}$ we can
efficiently obtain by iteration $I_m$ and $J_m$ for $m$ as high as required.
This iterative scheme has been implemented and verified, and is used
to give the island plots in this paper, which use simply $m=n/2$.

\section*{Acknowledgements}
I am grateful for useful discussions about transverse structure of
electron holes with Ivan Vasko. The codes that were used to do the
calculations and create the figures in this article are publically
available as doi:10.5281/zenodo.3746740 at
\url{https://zenodo.org/record/3746740} or at 
\url{https://github.com/ihutch/AxisymOrbits}. This work was
partially funded by NASA grant NNX16AG82G.

\bibliography{MyAll}

\end{document}